\documentclass[a4paper,fleqn,usenatbib]{mnras}

\usepackage{newtxtext,newtxmath}

\usepackage[T1]{fontenc}
\usepackage{ae,aecompl}

\usepackage{graphicx}	
\usepackage{amsmath}	
\usepackage{amssymb}	
\usepackage{multirow}
\usepackage{booktabs}
\usepackage{arydshln}

\title[Gaseous Environments of Quasars: AALs]
  {The Gaseous Environments of Quasars: Associate Absorption Lines with Density and Distance Constraints}
\author[C. Chen, F. Hamann, L. Simon and T. Barlow]
  {Chen Chen,$^{1,2}$\thanks{E-mail:chenomvmjs@ufl.edu}
 Fred Hamann,$^2$ Leah Simon$^{3,1}$ and Thomas Barlow$^4$\\
  $^1$Department of Astronomy, University of Florida, 211 Bryant Space Science Center, FL 32611, USA\\
$^2$Department of Physics \&\ Astronomy, University of California, Riverside, CA 92521, USA\\
$^3$Hands On Labs, 750 W. Hampden Ave., Suite 100, Englewood, CO 80110\\
$^4$Caltech Optical Observatories, 1200 E California Blvd, Pasadena, CA 91125, USA}

\usepackage{cleveref}

\newcommand{\cloudy}{\textsc{cloudy}}
\newcommand{\kms}{km s$^{-1}$}
\newcommand{\erg}{ergs s$^{-1}$}
\newcommand{\cmN}{cm$^{-2}$}
\newcommand{\cmn}{cm$^{-3}$}
\newcommand{\msun}{M$_{\odot}$}

\newcommand{\lam}{$\lambda$}

\newcommand{\lya}{\mbox{Ly$\alpha$}}
\newcommand{\lyb}{\mbox{Ly$\beta$}}
\newcommand{\lyc}{\mbox{Ly$\gamma$}}
\newcommand{\lyd}{\mbox{Ly$\delta$}}
\newcommand{\hi}{\mbox{H\,{\sc i}}}
\newcommand{\hii}{\mbox{H\,{\sc ii}}}

\newcommand{\civ}{\mbox{C\,{\sc iv}}}
\newcommand{\ciii}{\mbox{C\,{\sc iii}}}
\newcommand{\cii}{\mbox{C\,{\sc ii}}}
\newcommand{\siiv}{\mbox{Si\,{\sc iv}}}
\newcommand{\siiii}{\mbox{Si\,{\sc iii}}}
\newcommand{\siii}{\mbox{Si\,{\sc ii}}}
\newcommand{\nv}{\mbox{N\,{\sc v}}}
\newcommand{\niv}{\mbox{N\,{\sc iv}}}
\newcommand{\niii}{\mbox{N\,{\sc iii}}}

\newcommand{\ovi}{\mbox{O\,{\sc vi}}}

\newcommand{\oiv}{\mbox{O\,{\sc iv}}}
\newcommand{\oiii}{\mbox{O\,{\sc iii}}}

\newcommand{\feii}{\mbox{Fe\,{\sc ii}}}

\newcommand{\mgi}{\mbox{Mg\,{\sc i}}}
\newcommand{\mgii}{\mbox{Mg\,{\sc ii}}}

\newcommand{\neviii}{\mbox{Ne\,{\sc viii}}}

\date{  }

\pagerange{\pageref{firstpage}--\pageref{lastpage}} \pubyear{0000}

\def\LaTeX{L\kern-.36em\raise.3ex\hbox{a}\kern-.15em
    T\kern-.1667em\lower.7ex\hbox{E}\kern-.125emX}

\begin{document}

\label{firstpage}

\maketitle

\begin{abstract}

Associated absorption lines (AALs) in quasar spectra are valuable probes of the gas kinematics and physical conditions in quasar environments. The host galaxies are by definition in an active evolution stage that might involve large-scale blowouts and/or cold-mode accretion (infall) from the intergalactic medium (IGM). We discuss rest-frame UV spectra of four redshift 2--3 quasars selected to have low-ionisation AALs of \siii\ or \cii\ that place unique density and distance constraints on the absorbers. Our analysis of the AALs yields the following results. One of the quasars, Q0119$-$046, has a rich complex of 11 AAL systems that appear to be infalling at measured speeds from $\sim$0 to $\sim$1150 \kms\ at distance $\sim$5.7 kpc from the quasar. The range of ions detected, up to \neviii, indicates a wide range of densities from $\sim$4 to $\sim$2500 \cmn. Partial covering the quasar emission source requires cloud sizes $<$1 pc and possibly $<$0.01 pc. The short dissipation times of these small clouds suggests that they are created in situ at their observed location, perhaps as dense condensations in cold-mode accreting gas from IGM. The AALs in the other three quasars have outflow speeds from $\sim$1900 to $\sim$3000 \kms . Some of them also appear to have a range of densities based on the range of ions detected, including \mgi\ $\lambda$2853 in Q0105+061. However, the absence of excited-state AALs yields only upper limits on their gas densities ($\lesssim150$ \cmn) and large minimum distances ($\gtrsim$40 kpc) from the quasars. These AALs might represent highly extended quasar-driven outflows, although their actual physical relationships to the quasars cannot be established.  
\end{abstract}


\section{Introduction}

High-redshift quasars in the centers of massive galaxies signal an active early stage of massive galaxy evolution. Popular evolution models suggest that the black hole accretion that defines quasar activity is accompanied by a rapid burst of star formation in the host galaxies, triggered perhaps by a galaxy merger or interaction \citep{Sanders88, Elvis06, Hopkins08, Veilleux09, Hopkins16}. This activity is expected to continue until powerful outflows (feedback) driven by the central quasar and/or star formation lead to a blowout of gas and dust that quenches the starburst and cuts off the fuel supply for black hole accretion \citep{Silk98, Kauffmann00, King03, Scannapieco04, DiMatteo05, Ostriker10, Debuhr12, Rupke13, Rupke17, Cicone14}.

Infalling gas from the intergalactic medium (IGM, e.g., cold mode accretion) is also expected to be important during the early evolution stages to build galaxy mass, trigger star formation, and fuel the central black holes \citep{Katz03, Keres09, Keres12}. Recent observations show that massive gas reservoirs are indeed present around high-redshift quasars, and that they are more extended and more massive around quasar hosts than similar inactive (non-quasar) galaxies \citep[e.g.,][]{Prochaska14, Johnson15, Martin15, Martin16, Borisova16, Bouche16, Ho17}. These gas reservoirs are consistent with enhanced infall/cold-mode accretion from the IGM during the early/active quasars stages of massive galaxy evolution. It is likely that infall and outflow occur together if cold-mode accretion is involved in triggering the starbursts and quasars that also drive feedback \citep{Costa14, Nelson15, Suresh15}. 


Associated absorption lines (AALs) in quasar spectra are unique tools to study the gaseous environments of quasars and test models of massive galaxy evolution. AALs have narrow velocity widths (less than a few hundred \kms) much different from broad absorption lines (BALs), whose velocity widths $\gtrsim$2000 \kms\ clearly identify high-speed quasar-driven outflows \citep{Anderson87, Weymann91, Hamann04, Simon10, Muzahid13}. The term "associated" means they have absorption redshifts within several thousand \kms\ of the quasar emission-line redshift (i.e., $z_{abs}\approx z_{em}$, \citealt{Weymann79, Foltz86, Hamann97d}). It is known from statistical studies of large quasar samples that most AALs are intrinsic to the quasars, i.e., they form broadly within the environment of the quasar or its host galaxy \citep{Nestor08, Wild08, Perrotta16}. However, AALs can have a wide range of physical origins, from outflows near the quasars, to extended halos in the host galaxies, to cosmologically intervening gas or galaxies unrelated to the quasars \citep{Sargent82, Tripp96, Hamann01, Odorico04, Hamann11}. 


Several observational tests have been proposed to determine if individual AAL systems are likely intrinsic \citep[][]{Barlow97b, Hamann97e, Hamann99, Ganguly99, Wise04, Narayanan04, Misawa07c}. They include 1) high gas densities inferred from excited-state absorption lines (that require a close proximity to the quasars based on photoionisation constraints), 2) absorption line variability, 3) partial covering of the background light source that, for quasars, requires very small absorbing clouds, and 4) line profiles indicative of gas flows because they are too broad and smooth compared to thermal velocities \citep[see also][]{Hamann99, Srianand00, Schaye07, Ganguly99, Arav08}. These properties tend to go together. They are indicative of an intrinsic origin because they are more readily understood in terms of the dense dynamic of quasars compared to the larger, more quiescent, and lower density clouds expected for intervening absorption \citep{Rauch98}.

In this paper, we describe a study to understand the nature and origins of low-density associated absorbers that might reside in the extended host galaxies of quasars. This is an interesting subset of AALs because the extended galactic environments are where we might find evidence for cold mode accretion or quasar-driven winds directly interacting with the galactic interstellar medium. We describe the quasar sample and the data used in Section 2. We fit the AALs to measure the kinematics, column densities, and covering fractions in Section 3. We analyze the physical properties including ionisation, electron density, metallicity and radial distance in Section 4, and in Section 5 we discuss the origins and dynamics of absorption-line clouds. We conclude with a summary in Section 6. Throughout this paper, we adopt a cosmology with $H_0=71$ \kms\ Mpc $^{-1}$, $\Omega_M=0.27$ and $\Omega_{\Lambda}=0.73$.


\section{Data Overview}

\subsection{Quasar Sample}

We select four AAL quasars for our study from samples observed previously by our team using the High Resolution Echelle Spectrometer (HIRES) at the W. M. Keck Observatory (Keck) or the Utlraviolet and Visual Echelle Spectrometer (UVES) at the ESO Very Large Telescope (VLT). We require that the quasars have low-ionisation AALs of \cii\ or \siii\ detected in these existing high-resolution spectra. This sample is not intended to be representative of all quasar AALs. In particular, the low-ionization \cii\ and \siii\ lines are rare in AAL systems. However, these lines place unique constraints on the absorber densities and locations. They can, therefore, provide unique insights in the nature of the gas flows around quasars and their extended host galaxies. 

The quasars, their redshifts, and some basic information about the data are listed in \Cref{tab:sample}. Q0105+061, Q0334$-$204, and Q2044$-$168 are the only quasars with detected \cii\ or \siii\ AALs in the larger sample of 24 AAL quasars described by \citet{Simon12}. The 24 AAL quasars were selected from the Sloan Digital Sky Survey (SDSS) to be bright (apparent Magnitude $\lesssim$19) with \civ\ AALs at redshifts $\gtrsim$2. Q0119$-$046 is selected from unpublished high-resolution spectra obtained with both Keck and the Hubble Space Telescope (HST). Q0119$-$046 and Q2044$-$168 are radio-loud \citep{Murphy10, Chhetri13}, while Q0105+061 and Q0334$-$204 are radio-quiet \citep{Sramek80, Robson85}. The AALs in Q0119$-$046 were studied previously by \cite{Sargent82}. Here we present data with higher spectral resolution and higher signal-to-noise ratios, with wider wavelength coverage in the HST data that reveals additional absorption lines including \neviii\ \lam 770, 780, \ovi\ \lam 1032, 1038, and the Lyman lines to yield important new results on the nature of its AAL absorbers. 

Accurate systemic redshifts are important to judge infall versus outflow for the AALs and estimate the outflow kinetic energies. Redshifts derived from UV broad emission lines are known to be uncertain because the lines can be shifted in the quasar frame, with different lines shifted by different amounts. High-ionisation lines like \civ\ \lam 1548,1551 are typically blueshifted by several hundred \kms\ compared to narrow forbidden lines such as [\oiii] \lam 5007, which is generally regarded to be the best UV/optical indicator of quasar systemic redshifts \citep{Gaskell82, Shen07, Wang11}. Low-ionisation permitted lines such as \mgii\ \lam 2800 can also be good redshift indicators because their shifts relative to [\oiii] are typically $\sim$100 \kms\ \citep{Richards02}. We search the literature to find the best available redshift for each quasar based on the lines measured and the data quality. The results are listed in \Cref{tab:sample}. Below are some specific notes.

\begin{table}
\begin{minipage}{86mm}
\caption{Quasar data including the name, emission-line redshift, telescope-instrument used for observations, observation date (dd/mm/year), rest wavelength ranges, and spectral resolution $R=\lambda /\Delta\lambda$.}
\label{tab:sample}
\tabcolsep=0.12cm
\begin{tabular}{@{}cccccc}
\hline\hline
Quasar & $z_{\rm{em}}$ & Instrument & Obs. Date & $\lambda_{\rm{rest}}$ (\r{A}) & R\\
\hline
Q0105+061 & 1.960\footnote{\citet{Ulrich89}} & VLT-UVES & 09/22/2003 & 1184-1604 & 80000\\
\vspace{2mm}  &  &  &  & 2044-3352 & 110000\\
Q0119$-$046 & 1.9635\footnote{\citet{Steidel91}} & Keck-HIRES & 03/12/1996 & 1147-1989 & 45000\\
\vspace{2mm} &  & HST-FOS & 09/24/1996 & 750-1106 & 1300\\
Q0334$-$204 & 3.132\footnote{\citet{Tytler92}} & VLT-UVES & 09/23/2003 & 848-1077 & 80000\\
\vspace{2mm} &  &  &  & 1152-1654 & 110000\\
Q2044$-$168 & 1.939\footnote{\citet{Tytler04}} & VLT-UVES & 09/22/2003 & 1193-1606 & 80000\\
 &  &  &  & 2059-3376 & 110000\\
\hline
\end{tabular}
\end{minipage}
\end{table}

Q0105+061: \cite{Ulrich89} performed Gaussian fit (two-component fits for double lines) to measure the emission lines \civ, \ciii] \lam 1909, and \siiv+\oiv] \lam 1398 in the spectrum with resolution 7 \AA. The redshifts are 1.955, 1.953, 1.965, respectively, from the measurements of the three emission lines. We adopt $z_{em} = 1.960\pm 0.005$ ($\pm507$ \kms ) by weighting their individual line measurements based on the emission line widths.

Q0119$-$046: \citet{Steidel91} measured \mgii\ to determine the emission-line redshift $z_{em} = 1.9635$ by calculating the flux-weighted mean wavelength of the line above 85\% of its peak height. Given the spectral resolution and noise in the data, they estimated that the errors on their line positions are typically $\sim$50--100 \kms. We adopt the value $1.9635\pm0.0020$ ($\pm202$ \kms ), with a generous uncertainty to include the measurement uncertainty and the typical offset, $\sim$100 \kms, of \mgii\ relative to [\oiii] emission lines \citep{Richards02, Shen07}.

Q0334$-$204: \citet{Tytler92} presented a new maximum-likelihood method, which gave more accurate results than previous studies, to measure emission lines of \lya \lam 1216, \siiv+\oiv], \civ, and \ciii]. They obtained a value of $3.1322\pm 0.0013$ ($\pm94$ \kms ). We directly use this redshift.

Q2044$-$168: \citet{Tytler04} measured the emission-line redshifts using lines \lya, \nv\ \lam 1240, \siiv+\oiv], \civ, \ciii] with a typical error of 5 \AA\ for the peak wavelengths, and the redshifts determined by these lines are 1.939, 1.939, 1.936, 1.939, 1.938, respectively. We adopt the value of $1.939\pm 0.001$ ($\pm102$ \kms ) by weighting their individual line measurements.

\subsection{Observations and Data Reduction}

Basic information about the observations is listed in \Cref{tab:sample}. Spectra of Q0119$-$046 were observed with High Resolution Echelle Spectrometer (HIRES) on Keck and Faint Object Spectrograph (FOS) on the HST. The other three quasars were observed with UV-Visual Echelle Spectrograph (UVES) on the VLT. All of the spectra cover important absorption lines from at least \lya\ \lam 1216 to \civ\ \lam 1548, 1551 in the quasar rest frame. The HST-FOS spectrum of Q0119$-$046 provides additional UV coverage below the Lyman limit. The wavelength scales throughout this paper are vacuum heliocentric. We used standard techniques to reduce the Keck-HIRES and VLT-UVES spectra. In particular, for the Keck-HIRES spectra, we used a software package MAKEE \citep[as described in][]{Barlow97, Hamann97} for initial data reduction and spectral extraction. These procedures are described in detail in \citet{Hamann01}. The VLT-UVES data reduction procedures are described in \citet{simonphd}. 

We use the IRAF\footnote{Image Reduction and Analysis Facility (IRAF) is maintained and distributed by the National Optical Astronomy Observatories.} software package for additional data processing. In particular, we normalize the reduced Keck and VLT spectra to unity by fitting a pseudo-continuum to each quasar spectrum, including the emission lines. For regions with few absorption lines, we apply a polynomial fit locally around the absorption lines we will measure for our study. For the regions where the continuum is affected by significant absorption, e.g., in the \lya\ forest, we visually inspect the spectra to find small segments of continuum not affected by absorption or obvious noise spikes. We then interpolate between these segments, fitting the entire region with a low order polynomial.

\subsubsection{HST-FOS Spectrum of Q0119$-$046}

\begin{figure}
\centering
\includegraphics[width=84mm]{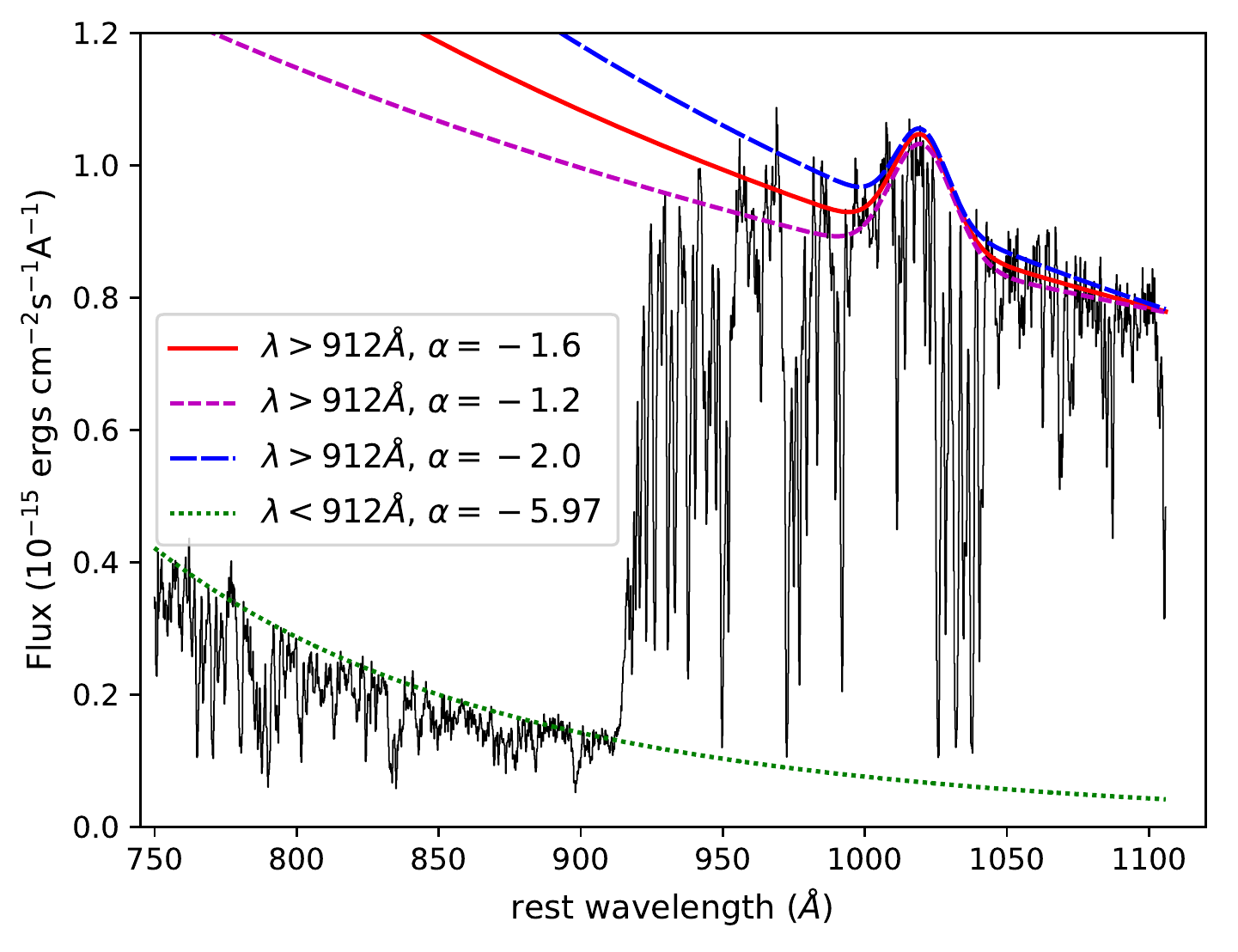}
\caption{HST spectrum of Q0119$-$046. The dotted green line shows the fit of the continuum at $\lambda<912$ \AA, and the solid red line shows the best estimation of the continuum at $\lambda>912$ \AA, including \ovi\ emission line. The dashed blue and mauve lines show upper and lower limits at $\lambda>912$ \AA, respectively.\label{fig:HST}}
\end{figure}

For Q0119$-$046 only, we include the UV spectrum obtained with HST-FOS shown in \Cref{fig:HST}. We retrieved this spectrum already reduced and calibrated from the Mikulski Archive for Space Telescopes (MAST). We attempted to test the wavelength calibrations using the Galactic absorption lines of \mgii\ \lam 2796, 2804 and \feii\ \lam 2587, 2600 \citep{Schneider93}. However, these lines are contaminated in the \lya\ forest and not useful. Instead, we compare the \lya\ AALs measured in the Keck spectrum (labeled as 2 and 7 in \Cref{fig:Q0119_2}) to the corresponding \lyb, \lyc\ and \lyd\ AALs in the HST spectrum (labeled as 2* and 7* in \Cref{fig:Q0119_2}). Our fits to these lines (described below) show no significant differences in the centroid redshifts and, therefore, we do not apply any corrections to the HST spectrum wavelengths. 

The HST spectrum has a much lower resolution than the Keck spectrum, but it covers shorter wavelengths with a variety of important AALs, including \neviii\ \lam 770, 780, \ovi\ \lam 1032, 1038, and the Lyman lines, plus strong Lyman limit absorption at the same redshift. \Cref{fig:HST} shows our estimates of the continuum on either side of the Lyman edge (912 \AA ). These continuum estimates are simple power laws of the form $F_{\lambda}\propto \lambda ^{\alpha}$, with a Gaussian profile added at $\sim$1034 \AA\ in the quasar frame to account for the broad \ovi\ \lam 1032, 1038 emission line. The AALs and numerous unrelated \lya\ forest lines make it difficult to locate the true continuum. We constrain our fits to using the median flux in narrow wavelength windows that appear relatively free of absorption lines. At wavelengths $>$912 \AA, the continuum slope near the Lyman edge is not well constrained due to absorption-line blending. We therefore consider three different continua (shown by the colored curves in \Cref{fig:HST}) that are upper and lower limits and a best-guess middle case based on visual inspection. 


\section{Absorption Features}

\subsection{Line Identifications}

\Cref{fig:Q0119_1,fig:Q0119_2,fig:Q0105,fig:Q0334,fig:Q2044} show the AAL profiles for all four quasars on a velocity scale relative to the emission-line redshifts. All of the quasars have multiple or complex/blended AAL systems. We identify individual AAL systems starting with the \civ\ \lam 1548, 1551 and \siiv\ \lam 1394, 1403 doublets because these lines are not contaminated in \lya\ forest and they tend to be strong and well-measured at wavelengths with high signal-to-noise ratios in our spectra. If these lines are broad or blended with neighboring features, we use weaker lines such as \cii\ \lam 1335 and \siii\ \lam 1260, 1304, and 1527, if available, to identify individual components and measure redshifts for the system. We then search other lines at these same redshifts. For Q0119$-$046, the search for additional lines includes the HST-FOS spectrum down to $\sim$750 \AA\ in the quasar rest frame. The lines identified in each quasar are listed in \Cref{tab:Q0119_1,tab:Q0119_2,tab:Q0105,tab:Q0334,tab:Q2044} below. We now describe the line measurements and the contents of these data tables.

\subsection{Line Profile Fits}

We fit each AAL with a Gaussian optical depth profile given by, 
\begin{equation}
\label{eq:1}
\tau_{v}=\tau_{0}e^{-v^2/b^2},
\end{equation}
where $\tau_v$ is the optical depth at velocity $v$, and $b$ is the Doppler parameter. $\tau_{0}$ is the line center optical depth, equal to
\begin{equation}
\label{eq:2}
\tau_{0}=\frac{\sqrt{\pi}e^2}{m_{e}c}\frac{Nf\lambda_{0}}{b},
\end{equation}
where $N$ is the column density, $f$ is the oscillator strength, and $\lambda_{0}$ is the laboratory line-center wavelength. We assume for simplicity that the background light source has a uniform brightness and the absorbing medium is homogeneous, with the same optical depth along every sightline, so the observed intensity at velocity is 
\begin{equation}
\label{eq:3}
\frac{I_{v}}{I_{0}}=1-C_v+C_ve^{-\tau_{v}},
\end{equation}
where $I_{0}$ is the continuum intensity, $I_{v}$ is the measured intensity at velocity $v$, $C_v$ is the covering fraction of the absorbing medium across the emission source, such that $0 < C_v \leq1$ \citep{Ganguly99, Hamann97b, Barlow97b}. It is known that actual absorbers in quasar spectra can be inhomogeneous, with a range of optical depths across the projected area of the emission source \citep{Barlow97b, deKool02, Hamann01, Hamann04, Arav02, Arav05}. In this situation, \Cref{eq:3} yields spatially-averaged optical depths and approximate covering fractions for material with $\tau_v \gtrsim 1$ \citep{Hamann04}. Real absorbers can also have velocity-dependent $C_v$ values that can differ between lines \citep{Barlow97, Hamann97b, Ganguly99, Hamann01, Hamann04}. The velocity-dependent effects are not important for the narrow absorption lines in our study. Therefore, for simplicity, we assume a constant covering fraction across the line profiles, e.g., $C_v=C_0$, but the value of $C_0$ can differ between lines. We also note that \Cref{eq:3} applies generically to partial covering situations regardless of its origins, which might include scattered light or partial covering of the accretion-disk continuum source versus the broad emission line region \citep{Ganguly99, Hamann04}. We discuss the origins of the partial covering further in Section 4.6.

To fit the lines in the low-resolution HST spectrum and the narrow AALs (velocity widths $\lesssim30$ \kms) in the Keck and VLT spectra, we convolve Gaussian optical depth profile with a Gaussian kernel that represents the instrumental broadening, specifically,
\begin{equation}
\label{eq:4}
\frac{I_{v}}{I_{0}}=\int e^{-\tau_{v'}}G(v-v')\: \mathrm{d}v',
\end{equation}
where $G(v)$ is Gaussian kernel defined by
\begin{equation}
\label{eq:5}
G(v)=\frac{1}{\sqrt{\pi}\Delta v_{ins}}e^{-\frac{\Delta v^2}{\Delta v_{ins}^2}},
\end{equation}
and $\Delta v_{ins}$ is the instrumental velocity resolution.

\begin{figure*}
\centering
\includegraphics[width=1.0\textwidth]{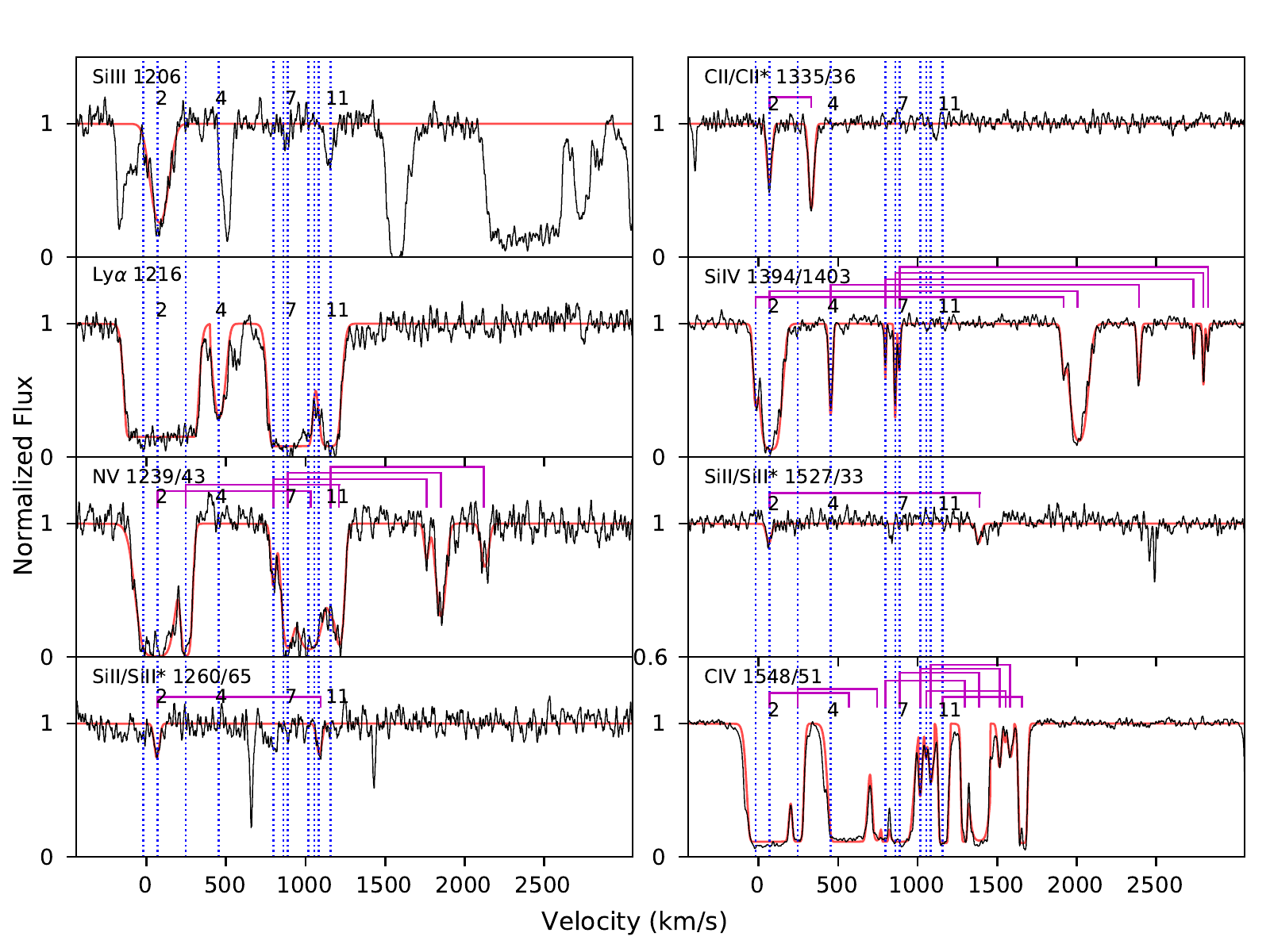}
\caption{Q0119$-$046: Normalized line profiles in the Keck-HIRES spectrum plotted on a velocity scale relative to the quasar redshift (\Cref{tab:sample}). The spectra are shown in black, and the fitting lines are shown in red. The blue dash lines are components from 1 to 11, and the brackets show the doublets. The velocities pertain to the short-wavelength lines in the doublets.\label{fig:Q0119_1}}
\end{figure*}

We derive estimates of $\tau_0$ and $C_0$ from \Cref{eq:3} by fitting the lines in doublets and multiplets simultaneously with fixed optical depth ratios determined by atomic physics. The lines are also required to have the same redshift, Doppler parameter, column density, and covering fraction. However, the tradeoff between $\tau_0$ and $C_0$ in the measured line depths can lead to ambiguous results in limiting cases where $\tau_0\gg 1$ or $\tau_0 \ll 1$, and where only a single line is available. Thus we consider four cases: 1) Heavily saturated lines based on the $\sim$1:1 observed depth ratio of the doublet and/or flat-bottom profiles that do not reach zero intensity, see for example components 2, 7 and 11 of \civ\ \lam 1548, 1551 and component 2 of \lya\ in \Cref{fig:Q0119_1}. In these situations, \Cref{eq:3} simplifies to $C_0\approx1-I_{v}/I_{0}$, where the covering fraction equals the observed depth of the line, and we adopt a conservative minimum optical depth of $\tau_0\gtrsim 3$ in the weakest doublet/multiplet component. 2) Doublets or multiplets that appear unsaturated based on intermediate line ratios, e.g., between 2:1 and 1:1 for the doublets\footnote{The optical depth ratios for lines sharing a common lower energy state are set by the ratio of their $f\lambda$ values, where $f$ is the oscillator strength and $\lambda$ is the line wavelength. This ratio is $\sim$2:1 for the doublets discussed in this paper, such as \civ\ $\tau(1548 A)/\tau(1551 A)$.}, such as component 2 of \siiv\ \lam 1394, 1403 in \Cref{fig:Q0119_1}). In this case, we use \Cref{eq:1,eq:3} to solve for both $C_0$ and $\tau_0$ by fitting the doublet lines simultaneously. Some examples of these fits and the partial covering determinations are described in Section 3.3.1 below (see also \Cref{fig:partial}). 3) Rare weak doublets that appear to have $\tau_0\ll 1$ based on $\sim$2:1 strength ratios, such as components 2 of \siii\ \lam 1260, 1527 in \Cref{fig:Q0119_1}. In this case, the values of $\tau_0$ and $C_0$ are degenerate in \Cref{eq:3} and cannot be determined separately. Thus we adopt $C_0=1$ to derive conservatively small lower limits on $\tau_0$ and the column densities. And 4) single lines that do not reach zero intensity, such as component 2 of \siiii\ \lam 1206 in \Cref{fig:Q0119_1}. Here again we adopt $C_0=1$ to derive conservatively small lower limits on the column densities. We find partial covering situations in components 2, 3, 5, 7, and 11 in Q0119$-$046, and component 2 in Q0334$-$204. For Q0119$-$046, we perform a detailed discussion in Section 3.3.1 below. And for Q0334$-$204, only component 2 in \civ\ shows partial covering, which is not convincing. We do not perform a further partial covering analysis on this quasar.

\begin{figure*}
\centering
\includegraphics[width=1.0\textwidth]{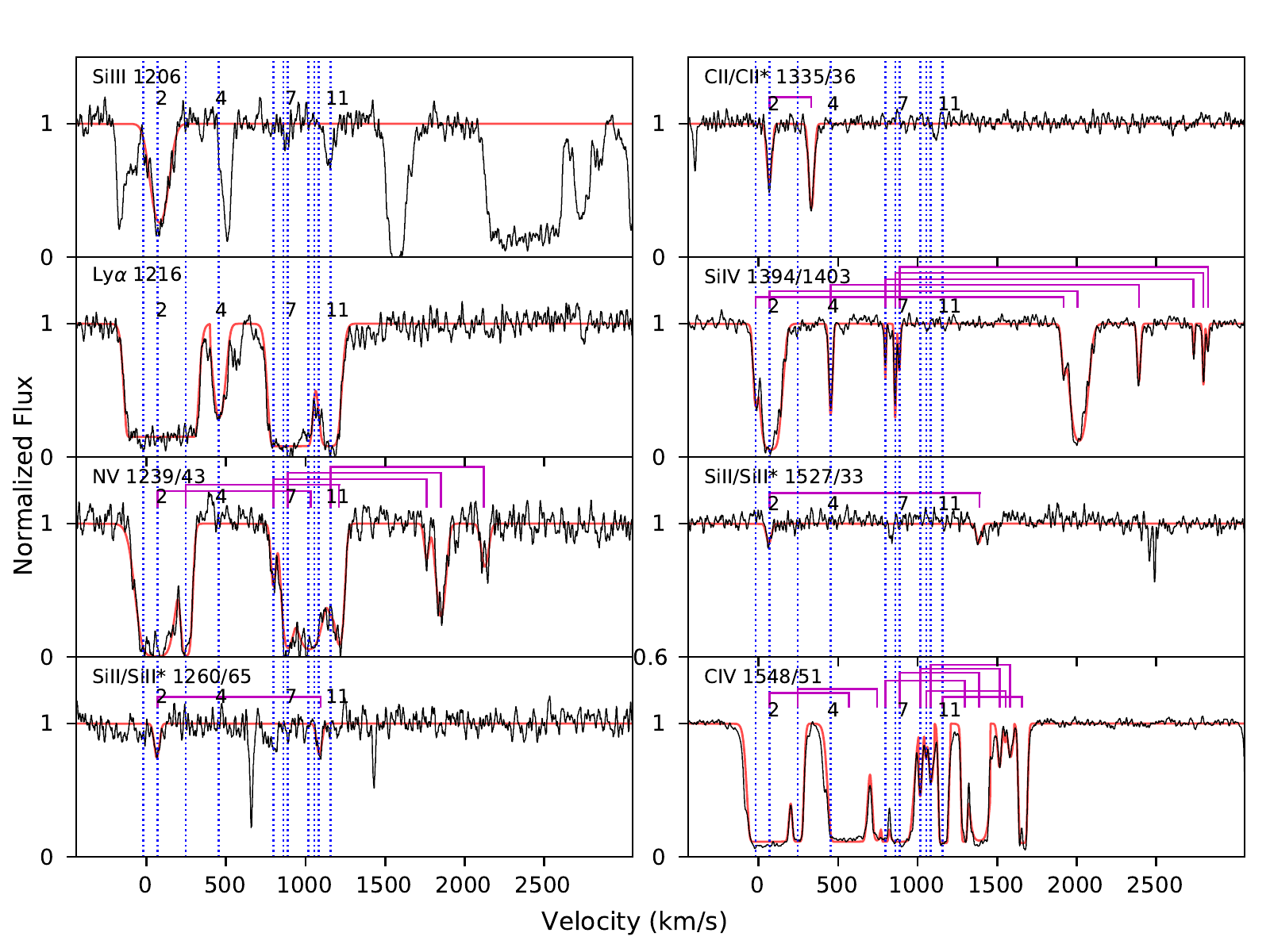}
\caption{Q0119$-$046: Normalized line profiles in the HST-FOS spectrum plotted on a velocity scale relative to the quasar redshift (\Cref{tab:sample}). The spectra are shown in black, and the fitting lines are shown in red. The blue dash lines are components, and the brackets show the doublets. The velocities pertain to the short-wavelength lines in the doublets. The bottom two panels are \lya\ and \nv\ spectra from Keck in comparison with the components of HST spectra.\label{fig:Q0119_2}}
\end{figure*}

\begin{figure*}
\centering
\includegraphics[width=1.0\textwidth]{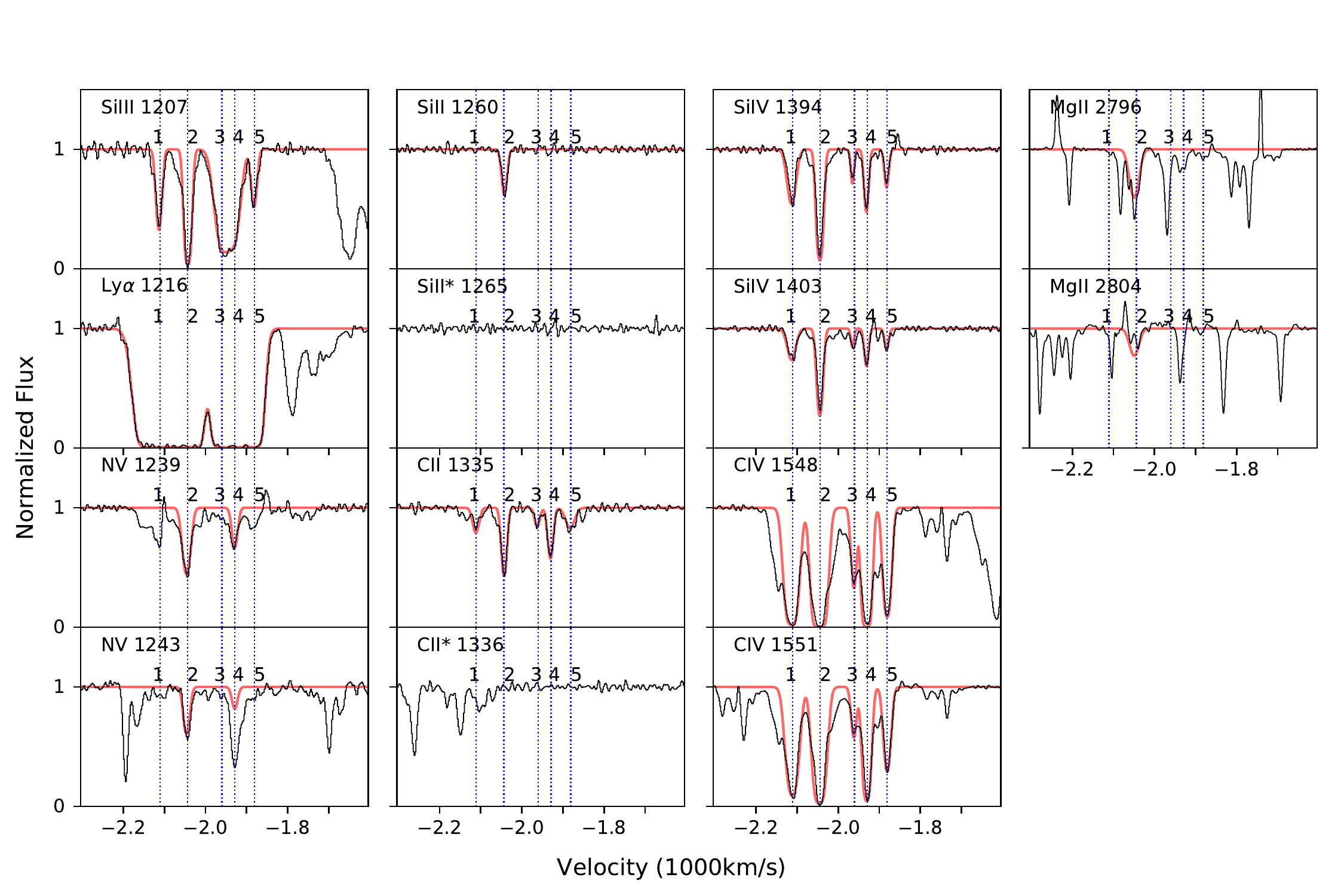}
\caption{Q2044$-$168: Normalized line profiles in the VLT-UVES spectrum plotted on a velocity scale relative to the quasar redshift (\Cref{tab:sample}). The spectra are shown in black, and the fitting lines are shown in red. The blue dash lines are components.\label{fig:Q2044}}
\end{figure*}

Another complication is line blending. If the blending is moderate between two or more components in the same transition, such that distinct components with separate minima are apparent (e.g., components 2 and 3 of \civ\ \lam 1548, 1551 in \Cref{fig:Q0119_1}), we fit all of the blended components simultaneously while allowing their $b$ values and centroids to be free parameters. If the blending is more severe, such that the observed feature forms a single line (e.g., components 1, 2 and 3 of \lya\ in \Cref{fig:Q0119_1}), we adopt $b$ values and centroid wavelengths determined from fits to other unblended lines of similar ionised ions in the same redshift system (if available), then fit the multiple lines in the blend simultaneously to derive their separate column densities and associated errors. Finally, if the blending involves unrelated lines (from different ion or features in the Lyman forest, e.g., component 7* of \niii\ \lam 990 in \Cref{fig:Q0119_2} and component 4 of \nv\ \lam 1243 in \Cref{fig:Q2044}), we again adopt $b$ values and centroid wavelengths determined from fits to other unblended lines of similar ionised ions in the same redshift system, and then fit the line to obtain a column density or maximum column density, depending of the severity of the blending. 

\begin{figure*}
\centering
\includegraphics[width=1.0\textwidth]{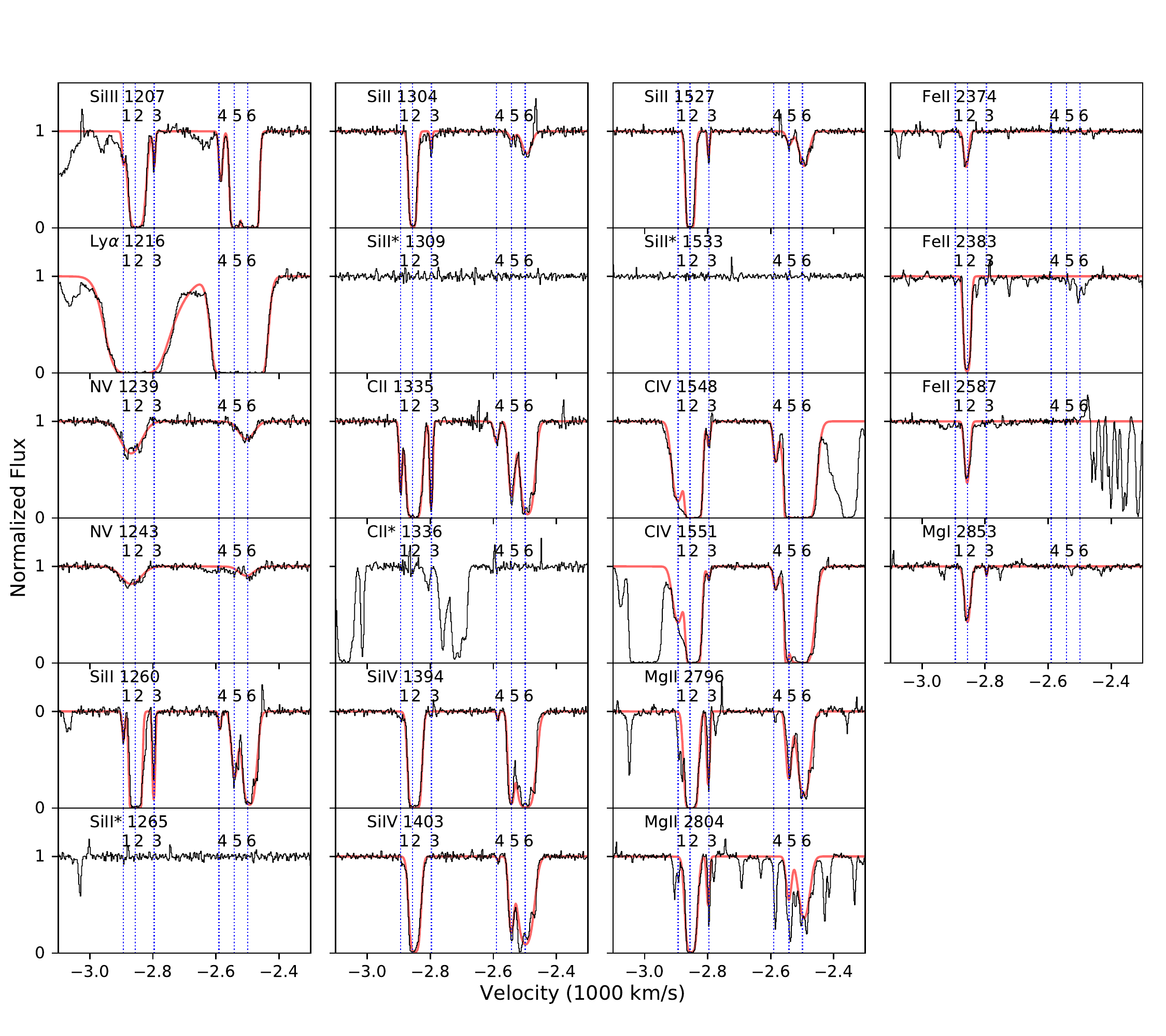}
\caption{Q0105+061: Normalized line profiles in the VLT-UVES spectrum plotted on a velocity scale relative to the quasar redshift (\Cref{tab:sample}). The spectra are shown in black, and the fitting lines are shown in red. The blue dash lines are components.\label{fig:Q0105}}
\end{figure*}

\begin{figure*}
\centering
\includegraphics[width=1.0\textwidth]{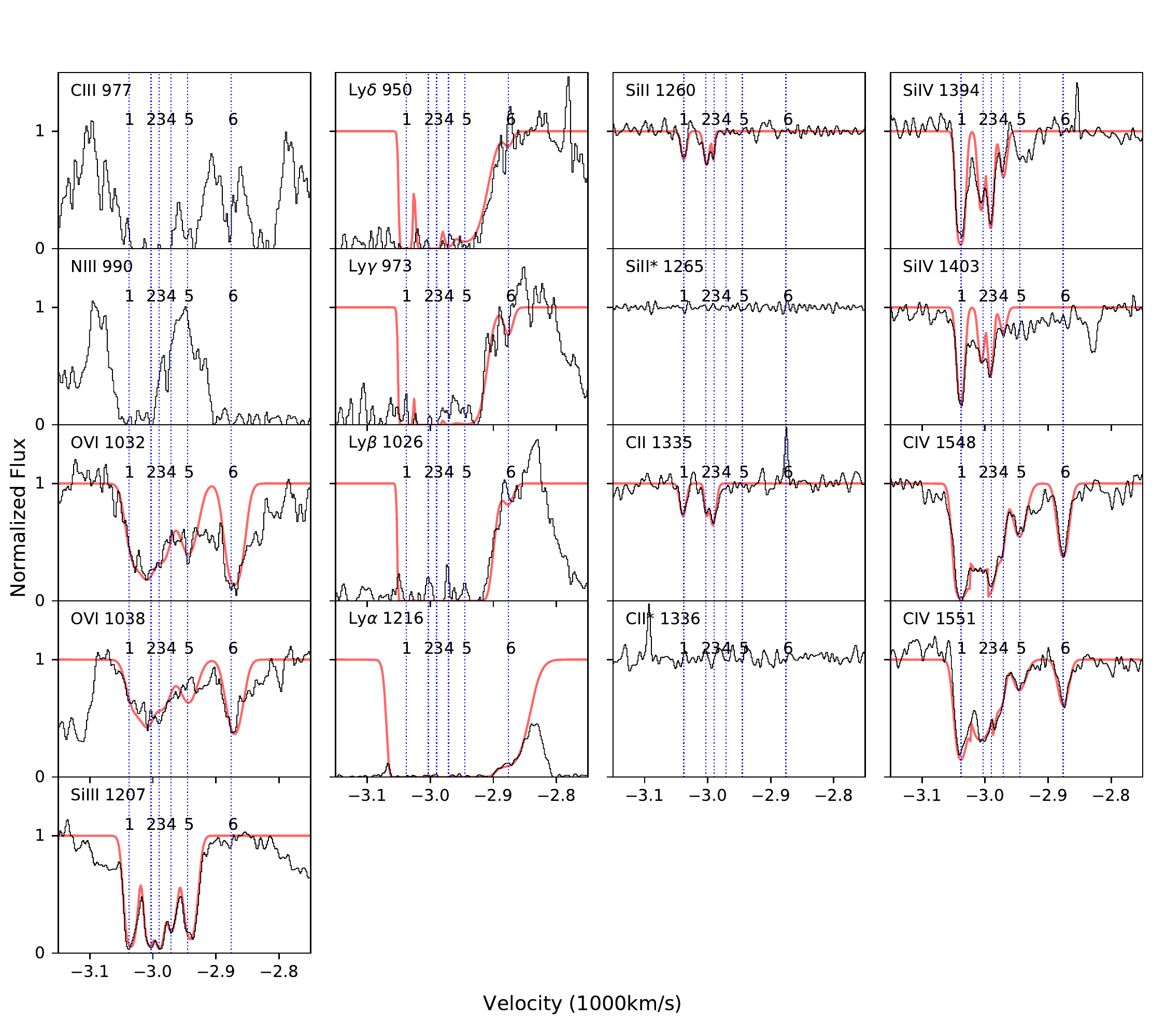}
\caption{Q0334$-$204: Normalized line profiles in the VLT-UVES spectrum plotted on a velocity scale relative to the quasar redshift (\Cref{tab:sample}). The spectra are shown in black, and the fitting lines are shown in red. The blue dash lines are components.\label{fig:Q0334}}
\end{figure*}

\Cref{tab:Q0119_1,tab:Q0119_2,tab:Q0105,tab:Q0334,tab:Q2044} list the derived line fit parameters and associated uncertainties. Footnotes in these tables indicate the procedure used for fitting the lines and deriving or assuming values of $C_0$, $\tau_0$, and $N$, while the Notes provide additional information on blends. The line data are organized in the tables according to the redshifts, i.e., component numbers in the first column as labeled in \Cref{fig:Q0119_1,fig:Q0119_2,fig:Q0105,fig:Q0334,fig:Q2044}. The uncertainties listed for most of the parameters are 1$\sigma$ errors derived from the fits, affected mainly by pixel-to-pixel noise fluctuations in the spectra. They do not consider errors in the continuum placement. For very optically thick lines based on $\sim$1:1 doublet ratios, we can obtain only lower limits on the column densities. We derive these limits from \Cref{eq:2} using $b$ values from our fits with $\tau_0\gtrsim 3$. If the saturated lines are blended with neighboring systems to make their $b$ values uncertain (e.g., components 1, 2 and 3 of \lya\ in \Cref{fig:Q0119_1}), we derive column density limits by combining $\tau_0\gtrsim 3$ with $b$ values measured from other unblended lines of similar ions in the same redshift system, as marked in the Notes in the tables. 

Finally, for the quasars Q0105+061, Q0334$-$204 and Q2044$-$168, we estimate upper limits on excited-state lines not detected, such as \siii* \lam 1533 and \cii* \lam 1336, if the corresponding resonance lines, e.g., \siii\ \lam 1527 and \cii\ \lam 1335, are cleanly measured. We use $b$ values and redshifts from the resonance lines to manually draw synthetic line profiles at the excited-state line positions and thereby determine $\sim$$3\sigma$ upper limits on the excited-state column densities. 

\begin{table*}
\centering
\begin{minipage}{140mm}
\caption{Individual absorption lines of Q0119$-$046 (Keck spectrum). Columns show component number, absorption redshift ($z_{abs}$) and the corresponding velocity shift ($v$) relative to the emission-line redshift, line identification and rest wavelength, observation wavelength, Doppler b parameter, logarithm of column density, covering fraction, and notes (sat=saturated line, bl=blended with neighboring systems, unbl=blended with unrelated lines (e.g., lines in the Lyman forest), w=weak line). For \lya\ lines, we obtain $b$ and $N(\hi)$ values from the fits of the Lyman limit and Lyman series, which is described in Section 3.3.2.}
\label{tab:Q0119_1}
\begin{tabular}{@{}cccccccc}
\hline \hline
$\#$ & $z_{abs}$ & ID & $\lambda_{obs}$ & $b$ & $\log N$ & $C_0$ & Notes\\ 
 & $v$ (\kms) & & (\AA) & (\kms) & (\cmN) & &\\
\hline
1 & 1.9633 & \siiv\ 1394 & 4130.37 & $15.3\pm 2.7$ & $13.11\pm0.06$ & 1.0$^b$ & \\
& $-20\pm202$ & \siiv\ 1403 & 4156.95 & ---& ---&  ---& \\ 
\hline
2 & 1.9642 & \siiii\ 1206 & 3576.41 & $64.3\pm8.1$ & $13.46\pm0.26$ & 1.0$^c$ & unbl? \\
& $71\pm202$ & \lya\ 1216 & 3603.74 & $80.8\pm10.3$ & $17.68\pm0.04$ & $0.85\pm0.02^a$ & bl \&\ sat\\
& & \nv\ 1239 & 3671.96 & $\sim53$ & $>14.7$ & 1.0$^a$ & bl \&\ sat, adopt $b$ from \siiv\ \\
& & \nv\ 1243 & 3683.76 & --- & --- & --- & ---\\
& & \siii\ 1260 & 3736.07 & $16.2\pm 2.1$ & $12.27\pm0.01$ & 1.0$^d$ & w\\
& & \siii*\ 1265 & 3749.71 & $16.2\pm2.1$ & $12.34\pm 0.01$ & 1.0$^d$ & w\\
& & \cii\ 1335 & 3955.75 & $18.8\pm 0.6$ & $13.66\pm 0.03$ & 1.0$^c$ & \\
& & \cii*\ 1336 & 3959.25 & $18.8\pm 0.6$ & $13.94\pm 0.02$ & 1.0$^c$ & \\
& & \siiv\ 1394 & 4131.69 & $52.5\pm 1.4$ & $14.40\pm 0.03$ & $0.95\pm 0.02^b$ & \\
& & \siiv\ 1403 & 4158.28 & --- & --- & --- & \\
& & \siii\ 1527 & 4525.40 & $16.2\pm 2.1$ & $12.27\pm0.01$ & 1.0$^d$ & w\\
& & \siii*\ 1533 & 4545.38 & $16.2\pm2.1$ & $12.34\pm 0.01$ & 1.0$^d$ & w\\
& & \civ\ 1548 & 4589.09 & $\sim53$ & $>14.6$ & $0.91\pm 0.01^a$ &  bl \&\ sat, adopt $b$ from \siiv\ \\
& & \civ\ 1551 & 4596.71 &  --- & --- & --- &  ---\\
\hline
3 & 1.9659 & \nv\ 1239 & 3674.29 & $29.2\pm 1.0$ & $14.68\pm 0.03$ & 1.0$^b$ & \\
& $243\pm202$ & \nv\ 1243 & 3686.11 &--- & --- & --- & \\
& & \civ\ 1548 & 4591.84 & $24.1\pm 0.6$ & $14.70\pm 0.10$ & $0.93\pm 0.02^a$ & sat\\
& & \civ\ 1551 & 4599.46 & --- & --- & --- & ---\\
\hline
4 & 1.9680 & \lya\ 1216 & 3608.12 & $42.9\pm 4.4$ & $13.66\pm 0.05$ & 1.0$^c$ & \\
& $456\pm202$ & \siiv\ 1394 & 4136.89 & $11.7\pm 1.7$ & $13.15\pm 0.05$ & 1.0$^b$ & \\
& & \siiv\ 1403 & 4163.52 & --- & --- & --- & \\
\hline
5 & 1.9714 & \nv\ 1239 & 3681.00 & $19.0\pm 1.8$ & $13.64\pm 0.03$ & 1.0$^b$ & \\
& $800\pm202$ & \nv\ 1243 & 3692.84 & --- & --- & ---& \\
& & \siiv\ 1394 & 4141.64 & $4.2\pm 1.7$ & $12.50\pm 0.11$ & 1.0$^d$ & w\\
& & \siiv\ 1403 & 4168.30 & --- & --- & --- & ---\\
& & \civ\ 1548 & 4600.25 & $11.4\pm 0.6$ & $14.74\pm 0.08$ & $0.92\pm 0.03^b$ & bl \\
& & \civ\ 1551 & 4607.88 & --- & --- & --- & \\
\hline
6 & 1.9720 & \siiv\ 1394 & 4142.49 & $4.8\pm 1.1$ & $12.92\pm 0.07$ & 1.0$^b$ & \\
& $860\pm202$ & \siiv\ 1403 & 4169.15 & --- & --- & --- & \\
\hline
7 & 1.9722 & \lya\ 1216 & 3613.52 & $66.5\pm10.5$ & $16.11\pm0.40$ & $0.91\pm0.02^a$ & bl \&\ sat \\
& $881\pm202$ & \nv\ 1239 & 3682.10 & $35.3\pm 1.3$ & $14.47\pm 0.02$ & 1.0$^b$ & bl\\
& & \nv\ 1243 & 3693.94 & --- & --- & --- & \\
& & \siiv\ 1394 & 4142.85 & $5.3\pm 2.5$ & $12.58\pm 0.11$ & 1.0$^d$ & w\\
& & \siiv\ 1403 & 4169.51 &--- & --- & --- & ---\\
& & \civ\ 1548 & 4601.71 & $\sim5$ & $>13.5$ & $0.93\pm 0.02^a$ &  bl \&\ sat, adopt $b$ from \siiv\ \\
& & \civ\ 1551 & 4609.35 & --- & --- & --- &  ---\\
\hline
\end{tabular}\\
$^a$ Heavily saturated lines, whose $C_0$ equals the observed depth of the line.\\
$^b$ Unsaturated doublets, where we solve for $C_0$ using \Cref{eq:3}.\\
$^c$ Unsaturated single lines, where we adopt $C_0=1$ because it is not constrained.\\
$^d$ Weak doublets, where we take a conservative approach by setting $C_0=1$.\\
\end{minipage}
\end{table*}

\begin{table*}
\centering
\begin{minipage}{140mm}
\contcaption{Individual absorption lines of Q0119$-$046 (Keck spectrum)}
\begin{tabular}{@{}cccccccc}
\hline \hline
$\#$ & $z_{abs}$ & ID & $\lambda_{obs}$ & $b$ & $\log N$ & $C_0$ & Notes \\
 & $v$ (\kms) & & (\AA) & (\kms) & (\cmN) & &\\
\hline
8 & 1.9736 & \civ\ 1548 & 4603.64 & $14.2\pm 5.8$ & $13.45\pm 0.13$ & 1.0$^b$ & \\
& $1022\pm202$ & \civ\ 1551 & 4611.28 & --- & --- & --- & \\
\hline
9 & 1.9739 & \civ\ 1548 & 4604.18 & $5.1\pm 3.8$ & $12.65\pm 0.29$ & 1.0$^d$ & w\\
& $1053\pm202$ & \civ\ 1551 & 4611.83 &--- & --- & --- &--- \\
\hline
10 & 1.9742 & \civ\ 1548 & 4604.71 & $19.6\pm 9.5$ & $13.43\pm 0.13$ & 1.0$^b$ & \\
& $1083\pm202$ & \civ\ 1551 & 4612.35 & --- & --- & --- & \\
\hline
11 & 1.9749 & \lya\ 1216 & 3616.47 & $\sim30$ & $>14.1$ & $0.91\pm0.02^a$ & bl \&\ sat, adopt $b$ from \nv\ \\
& $1154\pm202$ & \nv\ 1239 & 3685.50 & $29.7\pm 2.9$ & $13.91\pm 0.04$ & 1.0$^b$ &  bl \\
& & \nv\ 1243 & 3697.35 &  --- & --- & --- &  \\
& & \civ\ 1548 & 4605.90 & $\sim30$ & $>14.3$ & $0.90\pm 0.08^a$ & sat, adopt $b$ from \nv\ \\
& & \civ\ 1551 & 4613.54 & --- & --- & --- & ---\\
\hline
\end{tabular}
\end{minipage}
\end{table*}

\begin{table*}
\centering
\begin{minipage}{140mm}
\caption{Individual absorption lines of Q0119$-$046 (HST spectrum). See \Cref{tab:Q0119_1} for descriptions of the table contents. For the notes, nl=no obvious lines, bl=blended with neighboring systems, unbl=blended with unrelated lines (e.g., lines in the Lyman forest).}
\label{tab:Q0119_2}
\begin{tabular}{@{}cccccccc}
\hline \hline
$\#$ & $z_{abs}$ & ID & $\lambda_{obs}$ & $b$ & $\log N$ & $C_0$ & Notes \\
 & $v$ (\kms) & & (\AA) & (\kms) & (\cmN) & & \\
\hline
2* & 1.9646 & \niv\ 765 & 2268.60 & $117\pm28$ & $14.74\pm0.16$ & 1.0$^c$ & \\
& $111\pm202$ & \neviii\ 770 & 2284.05 & $147\pm18$ & $16.43\pm0.38$ & $0.68\pm0.07^b$ & \\
& & \neviii\ 780 & 2313.47 &--- & --- & --- & \\
& & \oiv\ 788 & 2335.84 & $110\pm19$ & $15.53\pm0.14$ &1.0$^c$ & \\ 
& & \lyc\ 973 & 2883.40 & $80.8\pm10.3$ & $17.68\pm0.04$ & $0.92\pm0.02^e$ & \\
& & \ciii\ 977 & 2896.04 & $111\pm22$ & $14.61\pm0.17$ & 1.0$^c$ &\\
& & \niii\ 990 & 2933.75 & $72.5\pm24.1$ & $14.08\pm0.03$ & 1.0$^c$ & \\
& & \lyb\ 1026 & 3040.69 & $80.8\pm10.3$ & $17.68\pm0.04$ & $0.92\pm0.02^e$ & \\
& & \ovi\ 1032 & 3059.38 & $172\pm41$ & $15.92\pm0.31$ & $0.90\pm0.08^b$ & \\
& & \ovi\ 1037 & 3076.33 &--- & --- & --- & \\
\hline
4* & 1.9680 & \niv\ 765 & 2270.53 & $98.9\pm44.1$ & $14.04\pm0.10$ & 1.0$^c$ & bl\\
& & \oiv\ 788 & 2337.89 & $101.9\pm51.0$ & $14.86\pm0.17$ &1.0$^c$ & bl \\ 
& & \lyc\ 973 & 2886.49 & $\sim43$ & $<14.8$ & $1.0^c$ & unbl, adopt $b$ from \lya\ \\
& & \ciii\ 977 & 2899.79 & $\sim100$ & $13.33\pm0.28$ & 1.0$^c$ & bl, adopt $b$ from \niv\ \\
& & \lyb\ 1026 & 3044.34 & $\sim43$ & $<14.8$ & $1.0^c$ & unbl, adopt $b$ from \lya\ \\
\hline
7* & 1.9725 & \niv\ 765 & 2275.25 & $48.9\pm16.4$ & $14.45\pm0.27$ & 1.0$^c$ & \\
& $911\pm202$ & \neviii\ 770 & 2290.08 & $\sim 50$ & $<15.0$ & 1.0$^c$ & bl, adopt $b$ from \ovi\ \\
& & \neviii\ 780 & --- & --- & --- & --- & nl\\
& & \oiv\ 788 & 2334.82 & $\sim 50$ & $<15.7$ & 1.0$^c$ & unbl, adopt $b$ from \niv\ \\
& & \lyc\ 973 & 2890.17 & $66.5\pm10.5$ & $16.11\pm0.40$ & 1.0$^b$ & \\
& & \ciii\ 977 & 2903.94 & $53.4\pm19.3$ & $14.34\pm0.49$ & 1.0$^c$ & \\
& & \niii\ 990 & 2940.37 & $27.5\pm11.5$ & $<14.5$ & 1.0$^c$ & unbl\\
& & \lyb\ 1026 & 3048.52 & $66.5\pm10.5$ & $16.11\pm0.40$ & 1.0$^b$ & \\
& & \ovi\ 1032 & 3067.21 & $49.4\pm15.4$ & $15.84\pm0.39$ & 1.0$^b$ & \\
& & \ovi\ 1037 & 3084.21 & --- & --- & ---& \\
\hline
11* & 1.9749 & \neviii\ 770 & 2291.95 & $\sim 45$ & $<14.4$ & 1.0$^c$ & bl, adopt $b$ from \ovi\ \\
& $1154\pm202$ & \neviii\ 780 & --- & --- & --- & --- & nl\\
& & \lyc\ 973 & 2893.20 & $\sim30$ & $14.71\pm0.32$ & $1.0^c$ & bl, adopt $b$ from \lya\ \\
& & \ciii\ 977 & 2906.53 & $\sim30$ & $13.68\pm0.31$ & 1.0$^c$ & bl, adopt $b$ from \lya\ \\
& & \lyb\ 1026 & 3051.42 & $\sim30$ & $14.71\pm0.32$ & $1.0^c$ & bl, adopt $b$ from \lya\ \\
& & \ovi\ 1032 & 3069.86 & $44.6\pm12.4$ & $15.18\pm0.52$ & 1.0$^b$ & \\
& & \ovi\ 1037 & 3086.88 &--- & --- & --- & \\
\hline
\end{tabular}\\
$^b$ Unsaturated doublets, where we solve for $C_0$ using \Cref{eq:3}.\\
$^c$ Unsaturated single lines, where we adopt $C_0=1$ because it is not constrained.\\
$^e$ We adopt $C_0$ and $N$ from fits of the Lyman limit, described in Section 3.3.2.
\end{minipage}
\end{table*}

\begin{table*}
\centering
\begin{minipage}{140mm}
\caption{Individual absorption lines of Q0105+061. See \Cref{tab:Q0119_1} for descriptions of the table contents. For the notes, nl=no obvious lines, sat=saturated line, bl=blended with neighboring systems, w=weak line, unbl=blended with unrelated lines (e.g., lines in the Lyman forest).}
\label{tab:Q0105}
\begin{tabular}{@{}cccccccc}
\hline \hline
$\#$ & $z_{abs}$ & ID & $\lambda_{obs}$ & $b$ & $\log N$ & $C_0$ & Notes\\ 
 & $v$ (\kms) & & (\AA) & (\kms) & (\cmN) & &\\
\hline
1 & 1.9315  & \siiii\ 1206 & 3536.79 & $\sim5$ & $11.90\pm0.33$ & 1.0$^c$ & adopt $b$ from \siii\ \\ 
& $-2894\pm507$ & \lya\ 1216 & 3563.67 & $\sim5$ & $>13.3$ & 1.0$^a$ & sat \&\ bl, adopt $b$ from \siii\ \\
& & \siii\ 1260 & 3694.86 & $5.4\pm2.9$ & $11.93\pm0.38$ & 1.0$^c$ & \\
& & \siii*\ 1265 & --- & --- & $<11.7$ & --- & nl \\
& & \cii\ 1335 & 3912.10 & $5.4\pm3.4$ & $13.42\pm0.23$ & 1.0$^c$ &  \\
& & \cii*\ 1336 & --- & --- & $<12.5$ & --- & nl \\
& & \civ\ 1548 & 4538.54 & $22.4\pm5.0$ & $14.04\pm0.10$ & 1.0$^b$ & bl \\
& & \civ\ 1551 & 4546.08 & --- & --- & --- & ---\\
\hline
2 & 1.9318 & \siiii\ 1206 & 3537.31 & $16.5\pm1.8$ & $14.0^{+1.0}_{-0.4}$ & 1.0$^c$ &\\
& $-2858\pm507$ & \lya\ 1216 & 3564.12 & $\sim15$ & $>13.8$ & 1.0$^a$ & sat \&\ bl, adopt $b$ from \siiv\ \\
& & \nv\ 1239 & 3631.84 & $42.5\pm18.6$ & $13.77\pm0.15$ & 1.0$^b$ & \\
& & \nv\ 1243 & 3643.51 &--- & --- & --- & \\
& & \siii\ 1260 & 3695.39 & $10.8\pm0.3$ & $14.43\pm0.04$ & 1.0$^b$ & \\
& & \siii*\ 1265 & --- & --- & $<12.6$ & --- & nl \\
& & \siii\ 1304 & 3824.24 & $10.8\pm0.3$ & $14.43\pm0.04$ & 1.0$^b$ & \\
& & \siii*\ 1309 & --- & --- & $<12.6$ & --- & nl\\
& & \cii\ 1335 & 3912.67 & $18.1\pm0.6$ & $14.7^{+0.7}_{-0.2}$ & 1.0$^a$ & sat \\
& & \cii*\ 1336 & --- & --- & $<12.8$ & --- & nl \\
& & \siiv\ 1394 & 4086.27 & $15.1\pm0.4$ & $14.1^{+0.7}_{-0.2}$ & 1.0$^b$ & \\
& & \siiv\ 1403 & 4112.70 & --- &--- & --- & --- \\
& & \siii\ 1527 & 4476.10 & $10.8\pm0.3$ & $14.43\pm0.04$ & 1.0$^b$ & \\
& & \siii*\ 1533 & -- & --- & $<12.6$ & --- & nl \\
& & \civ\ 1548 & 4538.93 & $\sim15$ & $14.7\pm0.5$ & 1.0$^a$ & bl, sat, adopt $b$ from \siiv\ \\
& & \civ\ 1551 & 4546.47 & --- & --- & --- & ---\\
& & \feii\ 2374 & 6961.38 & $9.0\pm1.0$ & $13.55\pm0.09$ & 1.0$^b$ & \\
& & \feii\ 2383 & 6985.73 & --- & --- & --- & \\
& & \feii\ 2587 & 7583.47 & --- & --- & --- & \\
& & \mgii\ 2796 & 8198.37 & $15.4\pm1.6$ & $13.77\pm0.07$ &1.0$^b$ & \\
& & \mgii\ 2804 & 8219.42 & --- & --- &--- & \\
& & \mgi\ 2853 & 8364.39 & $11.1\pm3.1$ & $12.08\pm0.10$ & 1.0$^c$ & \\
\hline
3 & 1.9324 & \siiii\ 1206 & 3537.96 & $4.3\pm2.2$ & $11.87\pm0.15$ & 1.0$^c$ &\\
& $-2797\pm507$ & \lya\ 1216 & 3564.84 & $\sim6$ & $>13.4$ & 1.0$^a$ & sat \&\ bl, adopt $b$ from \mgii\  \\
& & \siii\ 1260 & 3696.06 & $4.6\pm0.5$ & $12.68\pm0.02$ & 1.0$^b$ &\\
& & \siii*\ 1265 & --- & --- & $<12.7$ & --- & nl\\
& & \siii\ 1304 & 3824.94 & $4.6\pm0.5$ & $12.68\pm0.02$ & 1.0$^b$ &\\
& & \siii*\ 1309 & --- & --- & $<12.7$ & --- & nl\\
& & \cii\ 1335 & 3913.38 & $5.0\pm0.2$ & $13.64\pm0.03$ & 1.0$^c$ &\\
& & \cii*\ 1336 & --- & --- & $<12.3$ & --- & nl\\
& & \siii\ 1527 & 4476.91 & $4.6\pm0.5$ & $12.68\pm0.02$ & 1.0$^b$ &\\
& & \siii*\ 1533 & --- & --- & $<12.7$ & --- & nl\\
& & \civ\ 1548 & 4539.95 & $6.8\pm3.8$ & $12.66\pm0.18$ & 1.0$^d$ & w \\
& & \civ\ 1551 & 4547.48 & --- & --- & --- & ---\\
& & \mgii\ 2796 & 8199.99 & $5.7\pm0.7$ & $12.63\pm0.03$ & 1.0$^b$ &\\
& & \mgii\ 2804 & 8221.04 & --- & --- & --- &\\
& & \mgi\ 2853 & 8366.08 & $3.5\pm4.0$ & $10.67\pm0.49$ & 1.0$^c$ & w\\
\hline
\end{tabular}\\
$^a$ Heavily saturated lines, whose $C_0$ equals the observed depth of the line.\\
$^b$ Unsaturated doublets, where we solve for $C_0$ using \Cref{eq:3}.\\
$^c$ Unsaturated single lines, where we adopt $C_0=1$ because it is not constrained.\\
$^d$ Weak doublets, where we take a conservative approach by setting $C_0=1$.\\
\end{minipage}
\end{table*}

\begin{table*}
\centering
\begin{minipage}{140mm}
\contcaption{Individual absorption lines of Q0105+061}
\begin{tabular}{@{}cccccccc}
\hline \hline
$\#$ & $z_{abs}$ & ID & $\lambda_{obs}$ & $b$ & $\log N$ & $C_0$ & Notes\\ 
 & $v$ (\kms) & & (\AA) & (\kms) & (\cmN) & & \\
\hline
4 & 1.9345 & \siiii\ 1206 & 3540.47 & $5.7\pm3.8$ & $12.11\pm0.30$ & 1.0$^c$ &\\
& $-2590\pm507$ & \lya\ 1216 & 3567.31 & $\sim6$ & $>13.4$ & 1.0$^a$ & sat \&\ bl, adopt $b$ from \siii\ \\
&  & \siii\ 1260 & 3698.66 & $5.7\pm1.3$ & $11.72\pm0.49$ & 1.0$^c$ &\\
& & \siii*\ 1265 & --- & --- & $<11.5$ & --- & nl\\
& & \cii\ 1335 & 3916.10 & $10.1\pm1.2$ & $12.98\pm0.05$ & 1.0$^c$ &\\
& & \cii*\ 1336 & --- & --- & $<12.5$ & --- & nl\\
& & \siiv\ 1394 & 4089.98 & $4.8\pm2.3$ & $11.61\pm0.71$ & 1.0$^d$ & w\\
& & \siiv\ 1403 & 4116.42 & --- &--- & --- & --- \\
& & \civ\ 1548 & 4543.11 & $9.0\pm3.1$ & $13.03\pm0.14$ & 1.0$^b$ &  \\
& & \civ\ 1551 & 4550.64 & --- & --- & --- & \\
\hline
5 & 1.9349 & \siiii\ 1206 & 3541.02 & $\sim12$ & $>13.1$ & 1.0$^a$ & sat \&\ bl, adopt $b$ from \siii\ \\
& $-2544\pm507$ & \lya\ 1216 & 3567.88 & $\sim11$ & $>13.6$ & 1.0 & sat \&\ bl, adopt $b$ from \mgii\ \\
& & \siii\ 1260 & 3699.25 & $12.0\pm0.8$ & $12.77\pm0.05$ & 1.0$^b$ & \\
& & \siii*\ 1265 & --- & --- & $<12.0$ & --- &nl\\
& & \siii\ 1304 & 3828.24 & $12.0\pm0.8$ & $12.77\pm0.05$ & 1.0$^b$ &\\
& & \siii*\ 1309 & --- & --- & $<12.0$ & --- & nl\\
& & \cii\ 1335 & 3916.75 & $11.4\pm0.4$ & $13.82\pm0.02$ & 1.0$^c$ &\\
& & \cii*\ 1336 & --- & --- & $<12.3$ & ---& nl\\
& & \siiv\ 1394 & 4090.46 & $7.1\pm0.5$ & $13.45\pm0.05$ & 1.0$^b$ &\\
& & \siiv\ 1403 & 4116.92 & --- &--- & --- &\\
& & \siii\ 1527 & 4480.78 & $12.0\pm0.8$ & $12.77\pm0.05$ & 1.0$^b$ &\\
& & \siii*\ 1533 & --- & --- & $<12.0$ & ---& nl \\
& & \civ\ 1548 & 4543.73 & $\sim7$ & $>13.7$ & 1.0$^a$ & sat \&\ bl, adopt $b$ from \siiv\ \\
& & \civ\ 1551 & 4551.27 & --- & --- & --- & ---\\
& & \mgii\ 2796 & 8207.05 & $10.6\pm0.9$ & $12.85\pm0.1$ & 1.0$^c$ &\\
& & \mgii\ 2804 & 8228.12 &  --- & --- & --- & unbl\\
\hline
6 & 1.9353 & \siiii\ 1206 & 3541.57 & $\sim20$ & $>13.3$ & 1.0$^a$ & sat \&\ bl, adopt $b$ from \siii\ \\
& $-2503\pm507$ & \lya\ 1216 & 3568.41 & $\sim21$ & $>14.0$ & 1.0$^a$ & sat \&\ bl, adopt $b$ from \mgii\ \\
& & \nv\ 1239 & 3636.32 & $32.5\pm12.5$ & $13.36\pm0.06$ & 1.0$^b$ & \\
& & \nv\ 1243 & 3648.01 & --- & --- & --- & \\
& & \siii\ 1260 & 3699.83 & $19.7\pm0.6$ & $13.46\pm0.02$ & 1.0$^b$ &\\
& & \siii*\ 1265 & --- & --- & $<12.5$ & --- & nl\\
& & \siii\ 1304 & 3828.84 & $19.7\pm0.6$ & $13.46\pm0.02$ & 1.0$^b$ & \\
& & \siii*\ 1309 & --- & --- & $<12.5$ & ---& nl\\
& & \cii\ 1335 & 3917.41 & $20.6\pm0.5$ & $14.41\pm0.02$ & 1.0$^c$ &\\
& & \cii\* 1336 & --- & --- & $<12.3$ & --- & nl\\
& & \siiv\ 1394 & 4091.07 & $25.2\pm5.4$ & $14.0^{+0.7}_{-0.2}$ & 1.0$^b$ &\\
& & \siiv\ 1403 & 4117.53 & --- & --- & --- &\\
& & \siii\ 1527 & 4481.49 & $19.7\pm0.6$ & $13.46\pm0.02$ & 1.0$^b$ &\\
& & \siii*\ 1533 & --- & --- & $<12.5$ & --- & nl\\
& & \civ\ 1548 & 4544.40 & $\sim25$ & $>14.2$ & 1.0$^a$ & sat \&\ bl, adopt $b$ from \siiv\ \\
& & \civ\ 1551 & 4551.94 & --- & --- & --- & ---\\
& & \mgii\ 2796 & 8208.39 & $20.8\pm1.1$ & $13.31\pm0.03$ & 1.0$^b$ &\\
& & \mgii\ 2804 & 8229.47 &  --- & --- & --- &\\
\hline
\end{tabular}
\end{minipage}
\end{table*}

\begin{table*}
\centering
\begin{minipage}{140mm}
\caption{Individual absorption lines of Q0334$-$204. See \Cref{tab:Q0119_1} for descriptions of the table contents. For the notes, nl=no obvious lines, bl=blended with neighboring systems, sat=saturated line, w=weak line. For $N(\hi)$ lower limit estimates, we adopt $b$ from \civ, and assume $\tau_0>3$ in \lyd\ 950 \AA. We use \lyd\ because \lyd\ has smaller oscillator strength than \lya, \lyb\ and \lyc, which results in more accurate estimation of the lower limit.}
\label{tab:Q0334}
\begin{tabular}{@{}cccccccc}
\hline \hline
$\#$ & $z_{abs}$ & ID & $\lambda_{obs}$ & $b$ & $\log N$ & $C_0$ & Notes\\ 
 & $v$ (\kms) & & (\AA) & (\kms) & (\cmN) & & \\
\hline
1 & 3.0902 & \lyd\ 950 & 3884.64 & $\sim11$ & $>15.2$ & 1.0$^a$ & sat \&\ bl, adopt $b$ from \civ\ \\
& $-3035\pm94$ & \lyc\ 973 & 3977.87 & --- & --- & --- & ---\\
& & \lyb\ 1026 & 4195.35 & --- & --- & --- & --- \\
& & \lya\ 1216 & 4972.28 & --- & --- & --- & --- \\
& & \ovi\ 1032 & 4220.86 & $\sim11$ & $13.86\pm1.12$ & 1.0$^b$ & bl, adopt $b$ from \civ\ \\
& & \ovi\ 1038 & 4244.14 & --- & --- & --- & ---\\
& & \siiii\ 1206 & 4934.83 & $10.1\pm0.7$ & $13.02\pm0.04$ & 1.0$^c$ & \\
& & \siii\ 1260 & 5155.33 & $5.1\pm2.5$ &$ 11.84\pm0.22$ & 1.0$^c$ & \\
& & \siii*\ 1265 & --- & --- & $<11.2$ & --- & nl \\
& & \cii\ 1335 & 5458.43 & $6.3\pm5.5$ & $12.91\pm0.14$ & 1.0$^c$ & \\
& & \cii*\ 1336 & --- & --- & $<12.4$ & ---& nl\\
& & \siiv\ 1394 & 5700.66 & $6.8\pm0.3$ & $13.32\pm0.02$ & 1.0$^b$ & \\
& & \siiv\ 1403 & 5737.54 &--- & --- & --- & \\
& & \civ\ 1548 & 6332.35 & $10.6\pm0.9$ & $13.96\pm0.06$ & 1.0$^b$ & \\
& & \civ\ 1551 & 6342.86 & --- & --- & --- & \\
\hline
2 & 3.0906 & \lyd\ 950 & 3885.02 & $\sim16$ & $>15.4$ & 1.0$^a$ & sat \&\ bl, adopt $b$ from \civ\ \\
& $-3006\pm94$ & \lyc\ 973 & 3978.26 & --- & --- & --- & --- \\
& & \lyb\ 1026 & 4195.79 & --- & --- & --- & --- \\
& & \lya\ 1216 & 4972.92 & --- & --- & --- & --- \\
& & \ovi\ 1032 & 4221.17 & $\sim16$ & $14.05\pm1.34$ & 1.0$^b$ & bl, adopt $b$ from \civ\ \\
& & \ovi\ 1038 & 4244.45 & --- & --- & --- & ---\\
& & \siiii\ 1206 & 4935.32 & $9.0\pm0.9$ & $12.92\pm0.06$ & 1.0$^c$ & \\
& & \siii\ 1260 & 5155.95 & $5.7\pm1.5$ &$ 11.95\pm1.33$ & 1.0$^c$ & \\
& & \siii*\ 1265 & --- & --- & $<11.2$ & --- & nl\\
& & \cii\ 1335 & 5459.11 & $3.6\pm11.5$ & $12.65\pm0.33$ & 1.0$^c$ & \\
& & \cii*\ 1336 & --- & --- & $<11.9$ & ---& nl\\
& & \siiv\ 1394 & 5701.29 & $5.8\pm1.4$ & $12.96\pm0.20$ & 1.0$^b$ & \\
& & \siiv\ 1403 & 5738.17 & --- & --- & --- & \\
& & \civ\ 1548 & 6333.09 & $15.8\pm1.3$ & $14.4^{+0.4}_{-0.3}$ & $0.75\pm0.01^b$ & bl \\
& & \civ\ 1551 & 6343.60 & --- & --- & --- & \\
\hline
3 & 3.0908 & \lyd\ 950 & 3885.21 & $\sim9$ & $>15.1$ & 1.0$^a$ & sat \&\ bl, adopt $b$ from \civ\ \\
& $-2991\pm94$ & \lyc\ 973 & 3978.45 & --- & --- & --- & ---\\
& & \lyb\ 1026 & 4196.06 & --- & --- & --- & ---\\
& & \lya\ 1216 & 4973.26 & --- & --- & --- & ---\\
& & \ovi\ 1032 & 4221.53 & $\sim9$ & $13.83\pm1.74$ & 1.0$^b$ & bl, adopt $b$ from \civ\ \\
& & \ovi\ 1038 & 4244.80 & --- & --- & --- & ---\\
& & \siiii\ 1206 & 4935.49 & $7.2\pm1.3$ & $12.86\pm0.25$ & 1.0$^c$ & \\
& & \siii\ 1260 & 5156.14 & $2.1\pm1.2$ &$ 11.47\pm1.03$ & 1.0$^c$ &\\
& & \siii*\ 1265 & --- & --- & $<10.9$ & ---& nl\\
& & \cii\ 1335 & 5459.30 & $6.4\pm8.1$ & $13.04\pm0.04$ & 1.0$^c$ & \\
& & \cii*\ 1336 & --- & --- & $<12.4$ & --- & nl\\
& & \siiv\ 1394 & 5701.58 & $5.0\pm0.5$ & $12.94\pm0.03$ & 1.0$^b$ &\\
& & \siiv\ 1403 & 5738.46 & --- & --- & --- &\\
& & \civ\ 1548 & 6333.60 & $9.1\pm1.3$ & $13.48\pm0.06$ & 1.0$^b$ & \\
& & \civ\ 1551 & 6344.11 & --- & --- & --- & \\
\hline
\end{tabular}
$^a$ Heavily saturated lines, whose $C_0$ equals the observed depth of the line.\\
$^b$ Unsaturated doublets, where we solve for $C_0$ using \Cref{eq:3}.\\
$^c$ Unsaturated single lines, where we adopt $C_0=1$ because it is not constrained.\\
$^d$ Weak doublets, where we take a conservative approach by setting $C_0=1$.
\end{minipage}
\end{table*}

\begin{table*}
\centering
\begin{minipage}{140mm}
\contcaption{Individual absorption lines of Q0334$-$204}
\begin{tabular}{@{}cccccccc}
\hline \hline
$\#$ & $z_{abs}$ & ID & $\lambda_{obs}$ & $b$ & $\log N$ & $C_0$ & Notes\\ 
 & $v$ (\kms) & & (\AA) & (\kms) & (\cmN) & & \\
\hline
4 & 3.0911 & \lyd\ 950 & 3885.49 & $\sim8$ & $>15.1$ & 1.0$^a$ & sat \&\ bl, adopt $b$ from \civ\ \\
& $-2970\pm94$ & \lyc\ 973 & 3978.75 & --- & --- & --- & ---\\
& & \lyb\ 1026 & 4196.28 & --- & --- & --- & ---\\
& & \lya\ 1216 & 4973.36 & --- & --- & --- & ---\\
& & \ovi\ 1032 & 4221.70 & $\sim8$ & $13.13\pm1.21$ & 1.0$^b$ & bl, adopt $b$ from \civ\ \\
& & \ovi\ 1038 & 4244.98 & --- & --- & --- & ---\\
& & \siiii\ 1206 & 4935.89 & $10.1\pm2.9$ & $12.77\pm0.07$ & 1.0$^c$ & \\
& & \siiv\ 1394 & 5701.96 & $6.9\pm9.9$ & $12.50\pm2.00$ & 1.0$^d$ & w\\
& & \siiv\ 1403 & 5738.82 & --- & --- & --- & ---\\
& & \civ\ 1548 & 6333.75 & $7.8\pm0.8$ & $13.18\pm0.04$ & 1.0$^b$ & bl \\
& & \civ\ 1551 & 6344.27 & --- & --- & --- & --- \\
\hline
5 & 3.0914 & \lyd\ 950 & 3885.78 & $\sim14$ & $>15.3$ & 1.0$^a$ & sat \&\ bl, adopt $b$ from \civ\ \\
& $-2948\pm94$ & \lyc\ 973 & 3979.04 & --- & --- & --- & \\
& & \lyb\ 1026 & 4196.88 & --- & --- & --- & \\
& & \lya\ 1216 & 4973.87 & --- & --- & --- & \\
& & \ovi\ 1032 & 4222.11 & $17.7\pm16.6$ & $13.89\pm1.92$ & 1.0$^b$ & \\
& & \ovi\ 1038 & 4245.39 & --- & --- & --- & \\
& & \siiii\ 1206 & 4936.39 & $11.6\pm0.8$ & $12.90\pm0.04$ & 1.0$^c$ & \\
& & \civ\ 1548 & 6334.35 & $13.6\pm1.4$ & $13.28\pm0.03$ & 1.0$^b$ & \\
& & \civ\ 1551 & 6344.87 & --- & --- & --- & \\
\hline
6 & 3.0924 & \lyd\ 950 & 3886.73 & $10.5\pm3.7$ & $13.10\pm1.05$ & 1.0$^b$ & bl \\
& $-2875\pm94$ & \lyc\ 973 & 3980.01 & --- & --- & --- &--- \\
& & \lyb\ 1026 & 4197.72 & --- & --- & --- & ---\\
& & \lya\ 1216 & 4973.87 & --- & --- & --- & ---\\
& & \ovi\ 1032 & 4223.14 & $15.7\pm3.5$ & $14.19\pm0.56$ & 1.0$^b$ & \\
& & \ovi\ 1038 & 4246.42 & --- & --- & --- & \\
& & \civ\ 1548 & 6335.83 & $11.0\pm0.7$ & $13.40\pm0.02$ & 1.0$^b$ & \\
& & \civ\ 1551 & 6346.35 & --- & --- & --- & \\
\hline
\end{tabular}
\end{minipage}
\end{table*}

\begin{table*}
\centering
\begin{minipage}{130mm}
\caption{Individual absorption lines of Q2044$-$168. See \Cref{tab:Q0119_1} for descriptions of the table contents. For the notes, nl=no obvious lines, sat=saturated line, bl=blended with neighboring systems, unbl=blended with unrelated lines (e.g., lines in the Lyman forest).}
\label{tab:Q2044}
\begin{tabular}{@{}cccccccc}
\hline \hline
$\#$ & $z_{abs}$ & ID & $\lambda_{obs}$ & $b$ & $\log N$ & $C_0$ & Notes\\ 
 & $v$ (\kms) & & (\AA) & (\kms) & (\cmN) & & \\
\hline
1 & 1.9183 & \siiii\ 1206 & 3520.94 & $8.6\pm0.7$ & $12.51\pm0.03$ & 1.0$^c$ &\\
& $-2113\pm102$ & \lya\ 1216 & 3547.58 & $\sim17$ & $>13.8$ & 1.0$^a$ & sat \&\ bl, adopt $b$ from \civ\ \\
& & \cii\ 1335 & 3894.54 & $10.2\pm4.4$ & $13.0^{+0.1}_{-0.5}$ & 1.0$^c$ & \\
& & \cii*\ 1336 & --- & --- & $<12.7$ & --- & nl\\
& & \siiv\ 1394 & 4067.42 & $13.6\pm0.6$ & $12.89\pm0.01$ & 1.0$^b$ & \\
& & \siiv\ 1403 & 4093.71 & --- & --- & --- & \\
& & \civ\ 1548 & 4518.09 & $16.5\pm0.6$ & $14.35\pm0.04$ & 1.0$^b$ & \\
& & \civ\ 1551 & 4525.59 & --- & --- & --- & \\
\hline
2 & 1.9190 & \siiii\ 1206 & 3521.77 & $9.0\pm0.5$ & $13.01\pm0.04$ & 1.0$^c$ & \\
& $-2042\pm102$ & \lya\ 1216 & 3548.51 & $\sim16$ & $>13.8$& 1.0$^a$ & sat \&\ bl, adopt $b$ from \civ\  \\
& & \nv\ 1239 & 3616.07 & $12.2\pm1.7$ & $13.54\pm0.06$ & 1.0$^b$ & \\
& & \nv\ 1243 & 3627.70 & --- & --- & --- & \\
& & \siii\ 1260 & 3679.17 & $8.5\pm8.1$ & $12.26\pm0.27$ & 1.0$^c$ & \\
& & \siii*\ 1265 & --- & --- & $<11.2$ & --- & nl\\
& & \cii\ 1335 & 3895.49 & $8.2\pm0.4$ & $13.45\pm0.02$ & 1.0$^c$ & \\
& & \cii*\ 1336 & --- & --- & $<12.3$ & ---& nl\\
& & \siiv\ 1394 & 4068.34 & $9.0\pm0.3$ & $13.34\pm0.02$ & 1.0$^b$ & \\
& & \siiv\ 1403 & 4094.66 & --- & --- & --- & \\
& & \civ\ 1548 & 4519.15 & $15.8\pm0.7$ & $14.44\pm0.04$ & 1.0$^b$ &\\
& & \civ\ 1551 & 4526.65 & --- & --- & --- &\\
& & \mgii\ 2796 & 8162.36 & $15.2\pm5.9$ & $<12.49$ & 1.0$^b$ & unbl\\
& & \mgii\ 2804 & 8183.31 & --- & --- & --- & ---\\
\hline
3 & 1.9198 & \siiii\ 1206 & 3522.76 & $21.5\pm1.0$ & $13.10\pm0.04$ & 1.0$^c$ & bl\\
& $-1960\pm102$ & \lya\ 1216 & 3549.45 & $\sim8$ & $>13.5$ & 1.0$^a$ & sat \&\ bl, adopt $b$ from \civ\ \\
& & \cii\ 1335 & 3896.55 & $8.5\pm1.7$ & $12.76\pm0.01$ & 1.0$^c$ & \\
& & \cii*\ 1336 & --- & --- & $<12.1$ & ---& nl\\
& & \siiv\ 1394 & 4069.43 & $5.0\pm0.7$ & $12.22\pm0.03$ & 1.0$^b$ &\\
& & \siiv\ 1403 & 4095.75 &--- & --- & --- &\\
& & \civ\ 1548 & 4520.42 & $8.1\pm5.7$ & $13.33\pm0.04$ & 1.0$^b$& \\
& & \civ\ 1551 & 4527.92 & --- & --- & ---& \\
\hline
4 & 1.9201 & \siiii\ 1206 & 3523.11 & $18.0\pm1.1$ & $12.89\pm0.03$ & 1.0$^c$ & bl\\
& $-1929\pm102$ & \lya\ 1216 & 3549.87 & $\sim11$ & $>13.6$ & 1.0$^a$ & sat \&\ bl, adopt $b$ from \civ\ \\
& & \nv\ 1239 & 3617.50 & $9.1\pm7.0$ & $13.15\pm0.13$ & 1.0$^c$ & \\
& & \nv\ 1243 & 3629.12 & --- & --- & --- & unbl \\
& & \cii\ 1335 & 3896.97 & $8.5\pm0.6$ & $13.27\pm0.03$ & 1.0$^c$ & \\
& & \cii*\ 1336 & --- & --- & $<12.3$ & --- & nl\\
& & \siiv\ 1394 & 4069.89 & $7.0\pm0.4$ & $12.70\pm0.02$ & 1.0$^b$ & \\
& & \siiv\ 1403 & 4096.22 & --- & --- & --- & \\
& & \civ\ 1548 & 4520.89 & $10.7\pm0.6$ & $14.19\pm0.05$ & 1.0$^b$ &\\
& & \civ\ 1551 & 4528.40 & --- & --- & --- &\\
\hline
5 & 1.9206 & \siiii\ 1206 & 3523.66 & $9.0\pm0.9$ & $12.30\pm0.04$ & 1.0$^c$ & \\
& $-1878\pm102$ & \lya\ 1216 & 3550.44 & $\sim11$ & $>13.6$ & 1.0$^a$ & sat \&\ bl, adopt $b$ from \civ\  \\
& & \cii\ 1335 & 3897.59 & $14.8\pm2.1$ & $13.0^{+0.2}_{-0.3}$ & 1.0$^c$ &\\
& & \cii*\ 1336 & --- & --- & $<12.5$ & --- & nl\\
& & \siiv\ 1394 & 4070.56 & $6.7\pm0.6$ & $12.39\pm0.05$ & 1.0$^b$ &\\
& & \siiv\ 1403 & 4096.89 & --- & --- & --- &\\
& & \civ\ 1548 & 4521.65 & $10.6\pm0.5$ & $13.78\pm0.04$ & 1.0$^b$ &\\
& & \civ\ 1551 & 4529.16 & --- & --- & --- &\\
\hline
\end{tabular}
$^a$ Heavily saturated lines, whose $C_0$ equals the observed depth of the line.\\
$^b$ Unsaturated doublets, where we solve for $C_0$ using \Cref{eq:3}.\\
$^c$ Unsaturated single lines, where we adopt $C_0=1$ because it is not constrained.
\end{minipage}
\end{table*}

\subsection{Q0119$-$046}


Q0119$-$046 is a special case because it has a rich complex of blended AALs and additional wavelength coverage from HST that includes Lyman limit absorption and AALs measured at lower spectral resolution. 

\subsubsection{Absorption Lines}

We identify 11 distinct AAL redshift systems in the Keck spectrum (\Cref{fig:Q0119_1}), but only 4 systems in the HST spectrum due to the lower resolution and blending (\Cref{fig:Q0119_2}). We label the components in the HST spectrum 2*, 4*, 7*, and 11* to identify them loosely with the strongest components 2, 4, 7, and 11 in the Keck spectrum. \Cref{fig:Q0119_2} shows important AALs such as \ovi, \neviii\ and some Lyman lines in the HST spectrum compared to \civ\ and \nv\ measured in the Keck spectrum. The Keck spectra in \Cref{fig:Q0119_1} show that higher-ionisation lines (such as \civ) and stronger transitions (e.g., \lya) are broader and more blended. These lines also appear to be saturated based on $\sim$1:1 doublet ratios and/or flat-bottomed profiles that do not reach zero intensity. The depths of these saturated line yield the covering fractions (Section 3.2). 

\begin{figure}
\centering
\includegraphics[width=84mm]{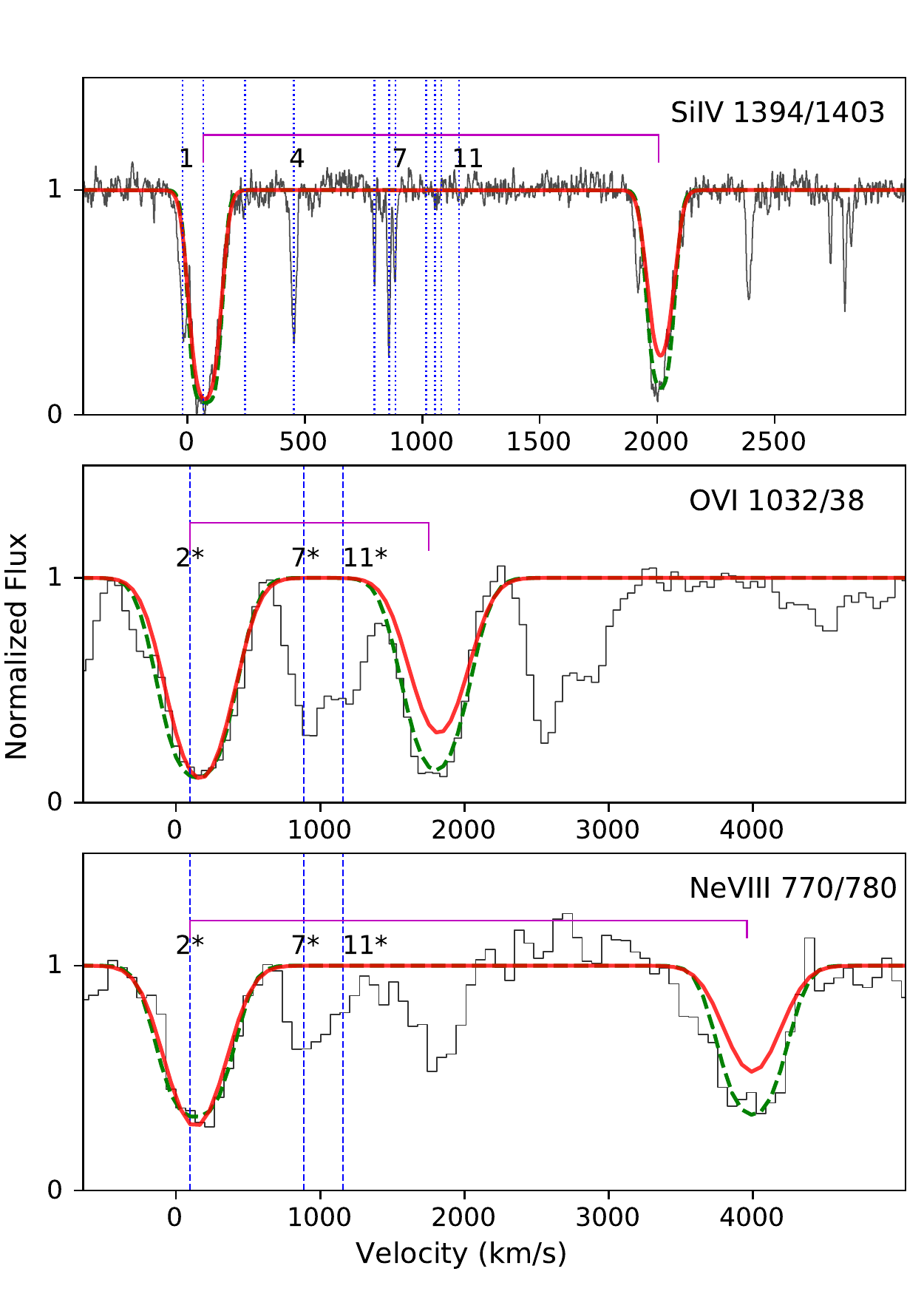}
\caption{Observed spectra of \siiv\ \lam 1394, 1403 (from Keck), \ovi\ \lam 1032, 1038, and \neviii\ \lam 770, 780 (from HST) of Q0119$-$046 (shown in black in each panel) are compared to predicted lines (component 2 or 2*) based on \siiv\ \lam 1394, \ovi\ \lam 1032 and \neviii\ \lam 770 assuming $C_0=1$ (red solid lines). The green dash lines in each panel show the final fitting results when $C_0<1$. The blue dash lines are components, and the brackets show the doublets. The velocities pertain to the short-wavelength lines in the doublets.\label{fig:partial}}
\end{figure}

Other lines require closer examination. \Cref{fig:partial} shows a partial covering analysis for components 2 and 2* in the doublets \siiv\ \lam 1394, 1403, \ovi\ \lam 1032, 1038 and \neviii\ \lam 770, 780. Specifically, fits to the stronger short-wavelength lines in these doublets are used to predict the strength of the weaker long-wavelength lines assuming $C_0 =1$. The predicted strengths are too weak in all cases, indicating $C_0 <1$ with specific $C_0$ values listed in \Cref{tab:Q0119_1,tab:Q0119_2}. There might be some ambiguity about these results for \ovi\ and \neviii\ because these lines are not well resolved in the HST spectrum. However, the comparisons to the well-resolved \nv\ lines in the Keck spectrum shown in \Cref{fig:Q0119_2}, and the tendency throughout the Q0119$-$046 spectrum for higher ion AALs to be broader and smoother (\Cref{fig:Q0119_1}) indicates that the \ovi\ and \neviii\ lines are reasonably resolved and {\it not} composed of narrow saturated features that could mimic partial covering at the HST spectral resolution. There is another possibility that the long-wavelength doublet components in \ovi\ and \neviii\ are made stronger to mimic partial covering due to blending with unrelated absorption line the \lya\ forest. However, there is no evidence for significant blending because the long-wavelength lines in both ions have redshifts and profiles as predicted from the short-wavelength lines. We conclude that significant partial covering does occur for \ovi\ and \neviii, with $C_0$ values given in \Cref{tab:Q0119_1,tab:Q0119_2}. 

\subsubsection{Lyman Limit and Lyman Series}

The Lyman limit in Q0119$-$046 provides an accurate measurement of the \hi\ column density and another estimate of the \hi\ covering fraction. We fit the Lyman limit absorption using \Cref{eq:3} with optical depths given by
\begin{equation}
\label{eq:6}
\tau_{\lambda}=6.63\times 10^{-18}N(\hi) \left( \frac{\lambda}{912{\rm \AA}} \right) ^3,
\end{equation}
where $N(\hi)$ is the \hi\ column density in cm$^{-2}$ \citep{Osterbrock89}. The main uncertainty in this analysis is the unabsorbed continuum intensity, $I_0$. For this we extrapolate our best-fit power law at $\lambda>912$ \AA\ with $\alpha=-1.6$ (Section 2.2.1 and \Cref{fig:HST}) to shorter wavelengths, and we use the range of possible values from $\alpha = -1.2$ to $-2.0$ to assess the uncertainties. We also fix the redshift to that of component 2* (\Cref{tab:Q0119_2}) because this system includes many strong Lyman lines and it is the only system with an \hi\ column density large enough to produce a Lyman limit (based on our absorption-line assessments). We divide by this power law and then fit the resulting normalized continuum at $\lambda < 912$ \AA\ using \Cref{eq:3,eq:6}, constrained by the median flux in narrow wavelength intervals that appear unaffected by \lya\ forest lines. 

\begin{figure}
\centering
\includegraphics[width=85mm]{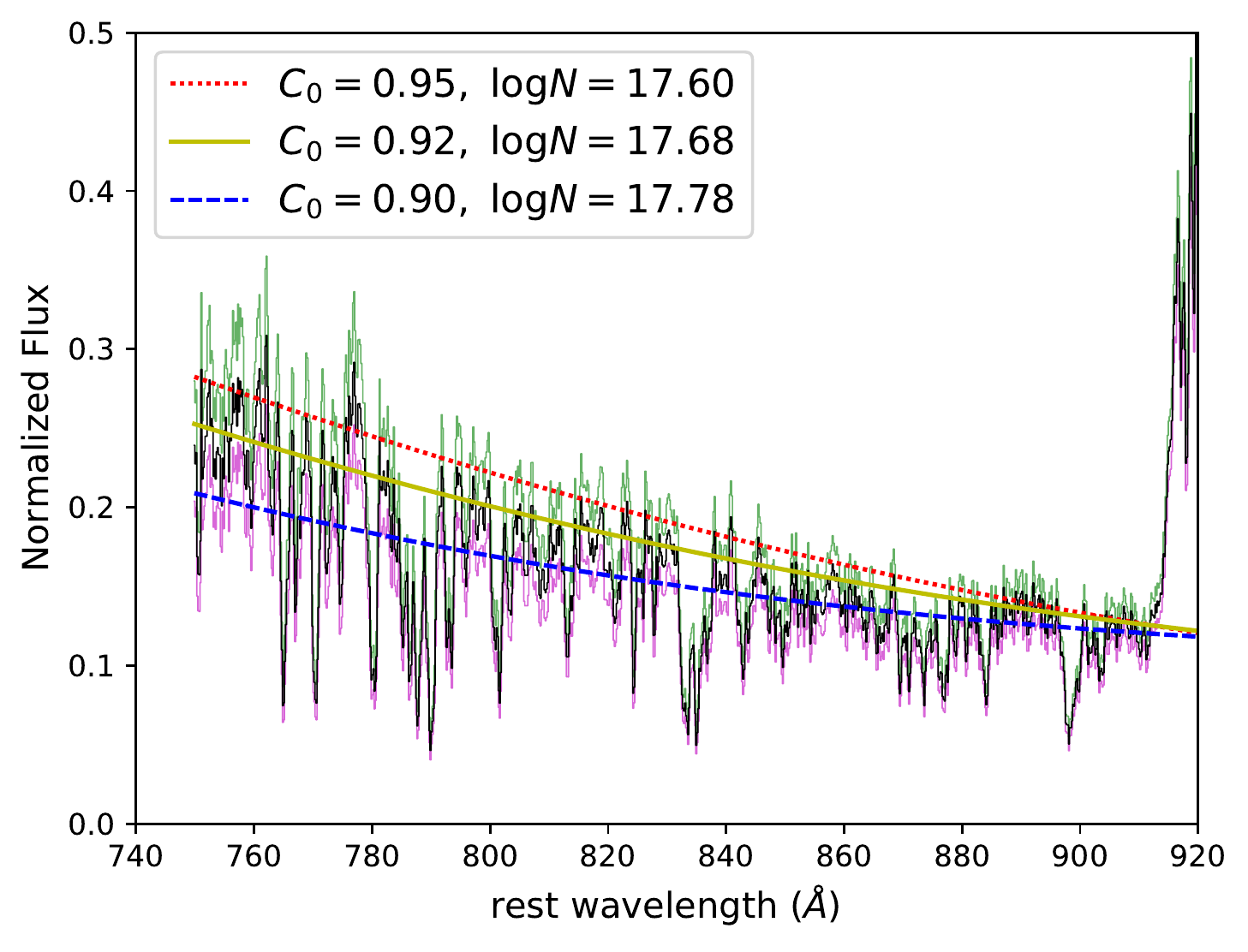}
\caption{Observed spectrum normalized in three different ways using the continuum fits from Section 2.2.1 (\Cref{fig:HST}). Column densities values change with different covering fractions by fitting the Lyman limit. The green and mauve spectra are the upper and lower limits of the normalized spectra, respectively. And the black spectrum shows the best normalized spectrum. The dotted red and dashed dark blue lines are the upper and lower limits of fits considering uncertainties of normalization. The range of $C_0$ is 0.90 ($\log N=17.78$) to 0.95 ($\log N=17.60$). The best fit is $C_0=0.92$ and $\log N=17.68$, shown as the solid yellow line.\label{fig:limit_fit}}
\end{figure}

The results (shown in \Cref{fig:limit_fit}) indicate a covering fraction $C_0=0.92\pm0.02$ and column density $\log N(\hi)({\rm cm}^{-2})=17.68\pm0.04$ based our best-fit continuum with $\alpha = -1.6$. Additional fits using the full range of continua from \Cref{fig:HST} ($\alpha = -1.2$ to $-$2.0) yield firm lower and upper limits $C_0=0.90$ ($\log N(\hi)({\rm cm}^{-2})=17.78$) and $C_0=0.95$ ($\log N(\hi)({\rm cm}^{-2})=17.60$), respectively. These covering fractions are consistent with the upper lines in the Lyman series (component 2* in \Cref{fig:Q0119_2}) but slightly larger than the well-determined value of $C_0=0.85 $ for \lya\ (\Cref{fig:Q0119_1}, \Cref{tab:Q0119_1}). We attribute the anomaly in \lya\ to its location on top of strong and broad \lya\ emission line, indicating that the \lya\ absorber only partially covers the broad emission-line region (see Section 4.6 for more discussion). We use a Gaussian function to model the \lya\ emission line and try to eliminate its effect. But our model is only a lower limit due to the severe contamination by the strong absorptions lines. This lead to a smaller value of $C_0$ for \lya.

These results for $C_0$ and $\log N(\hi)$ from the Lyman limit are more reliable than what can be determined from Lyman series lines because the lines are under-resolved in the HST spectrum and they are contaminated by the \lya\ forest. Therefore, we adopt $C_0=0.92$ and $\log N(\hi)({\rm cm}^{-2}) =17.68$ and apply our fitting procedure to the upper Lyman series lines to determine that they have $b=81\pm 10$ \kms\ (see component 2* in \Cref{fig:Q0119_2}). 

\section{Analysis}

In this section, we analyze the AAL systems that have resonance lines of \cii\ or \siii\ detected so that our measurements or upper limits on the corresponding excited-state lines, \cii* or \siii*, provide density and location constraints on the absorption environments. This includes component 2 in Q0119$-$046, all components in Q0105+061, components 1, 2 and 3 in Q0334$-$204, and all components in Q2044$-$168. These AAL systems also include multiple ions of the same element, such as \cii/\civ\ and \siii/\siiii/\siiv, that we use to estimate the ionisations, total column densities, and element abundances compared to photoionisation models. \Cref{tab:para} lists various derived parameters for these systems, which we discuss below. 

\subsection{Electron Number Density}

\cii\ and \siii\ have similar energy level structures with a ground $^2$P$^o$ term that is split into a true ground $J=1/2$ state and a slightly excited $J=3/2$ state. This leads to pairs of \cii*/\cii\ and \siii*/\siii\ lines whose ratios are density dependent because the excited $J=3/2$ states are populated mainly by collisions from the $J=1/2$ ground \citep{Bahcall68, Sargent82, Morris86, Hamann01, Dunn10, Arav13, Finn14}. For a simple two-level atom where the excited-state energy is small compared to $kT$, the density dependence can be written as
\begin{equation}
\label{eq:7}
n_e = n_{cr}\bigg(\frac{N_{lo}}{N_{up}}\frac{g_{up}}{g_{lo}}-1\bigg)^{-1}
\end{equation}
where $n_e$ is the electron density, $n_{cr}$ is the critical electron density of the upper state, $N_{lo}$ and $N_{up}$ are the column densities in the ground and excited states, respectively, and $g_{up}/g_{lo} =2$ is the ratio of their statistical weights. For a temperature of $10^4$ K, we derive $n_{cr}\approx1766$ \cmn\ for the \siii\ upper state and $n_{cr}\approx49$ \cmn\ for \cii\ (using collision strengths from \citet{Tayal08a, Tayal08b} and radiative decay rates from the NIST atomic spectra database\footnote{http://www.nist.gov/pml/data/asd.cfm}). 

\Cref{tab:para} shows the electron densities in \siii\ or \cii\ regions that result from \Cref{eq:7} combined with the column density data in \Cref{tab:Q0119_1,tab:Q0119_2,tab:Q0105,tab:Q0334,tab:Q2044}. The excited-state lines are detected only in component 2 of Q0119$-$046. The density listed for Q0119$-$046 derives from the $N(\siii*)/N(\siii)$ with the result $\log n_e$ (\cmn) $= 3.4\pm0.3$. This is consistent with the observed ratio of $N(\cii*)/N(\cii)$ near 2, which yields only a lower limit $\log n_e$ (\cmn) $\gtrsim 2.2$ \citep[consistent with][]{Sargent82}. 

All of the other quasars with non-detections provide only upper limits on the densities. The upper limits listed for components 2, 3, 5 and 6 in Q0105+061 are based on $N(\siii*)/N(\siii)$ because we fit multiple \siii\ lines (1260, 1304, and 1527 \AA) simultaneously with the same $b$ parameter, velocity shift and column density, which lead to reasonably good constraints. The upper limits listed for components 1 and 4 in Q0105+061 are based on $N(\cii*)/N(\cii)$ because \siii\ lines are weak and only \siii\ \lam 1260 is detected. The densities listed for Q0334$-$204 and Q2044$-$168 derive from $N(\cii*)/N(\cii)$ because the resonance \cii\ \lam 1335 lines are well measured and this ratio provides smaller upper limits than $N(\siii*)/N(\siii)$. The upper limits in these quasars range from $\lesssim$150 to $\lesssim$15 \cmn.

\subsection{Photoionisation Models}

We assume that the absorbers are in photoionisation equilibrium with the quasar radiation field and examine their ionisation properties using the computer code \cloudy\ (version 13.03, \citeauthor{Ferland13} 2013). The calculations use a plane-parallel absorbing geometry, solar abundances, and a constant gas density $n_{\rm{H}}=100$ \cmn. The specific value of the density is not important for our analysis of the ground-state column densities \citep{Hamann97d, Hamann02, Leighly11}. Recent work by \citet{Baskin14} favors constant pressure absorbing clouds instead of constant density for broad absorption line outflows. However, experiments with \cloudy\ show that this also does not have an impact on our results. The ionisation structure does depend on the shape and intensity of the incident spectral energy distribution (SED). We use an SED that is roughly characteristic of luminous quasars \citep{Reeves00, Richards06, Hopkins07, Shull12}. It is described by a power law, $f_{\nu}\propto\nu^{\alpha}$, with spectral indices $\alpha_{UV}=-0.5$ and $\alpha_{X}=-0.9$ at UV and X-ray wavelengths, respectively. These segments are normalized to yield the two-point power law index $\alpha_{OX}=-1.7$ between 2500 \AA\ and 2 keV \citep{Hamann13}. The UV and X-ray segments connect smoothly using a Planck exponential with temperature 350,000 K. We specify the intensity of this radiation field using the ionisation parameter, $U$, which is defined as the dimensionless ratio between the number density of hydrogen-ionising photons at the illuminated face of the clouds to the number density of hydrogen atoms, e.g.,
\begin{equation}
\label{eq:8}
U\equiv \frac{\Phi(\mbox{H})}{n_{\rm{H}}c},
\end{equation}
where $n_{\rm{H}}$ is the total hydrogen density, $c$ is the speed of light, and $\Phi(\mbox{H})$ is the flux of H-ionising photons given by
\begin{equation}
\label{eq:9}
\Phi(\mbox{H})=\frac{1}{4\pi R^2}\int_{\nu_1}^{\infty} \frac{L_{\nu}}{h\nu} \, \mathrm{d}\nu,
\end{equation}
where $L_{\nu}$ is the quasar luminosity density, $R$ is the radial distance between the absorber and the quasar, and $h\nu_1=13.6$ eV. 

We run the \cloudy\ simulations with a fixed \hi\ column density $\log N(\hi)$(\cmN)~$=17.7$ and a total column density $\log N$(H)(\cmN)~$=20.9$ for Q0119$-$046 based on our measurements in Section 3.3.2 (\Cref{tab:Q0119_1,tab:Q0119_2}) and Section 4.3 below. For the other three quasars, we adopt $\log N(\hi)$(\cmN)~$=16$ consistent with the lower limits listed in \Cref{tab:Q0105,tab:Q0334,tab:Q2044}. We adopt $\log N(\hi)$(\cmN)~$=16$ to illustrate the results in a regime where the clouds are optically thin in the Lyman continuum such that the calculated ionisations do not depend on the specific value of $N(\hi)$. The actual \hi\ column densities are not known for these quasars. We discuss other possibilities in Section 4.3 below.

\subsection{Ionisation \& Total Column Densities}

\begin{figure}
\centering
\includegraphics[width=84mm]{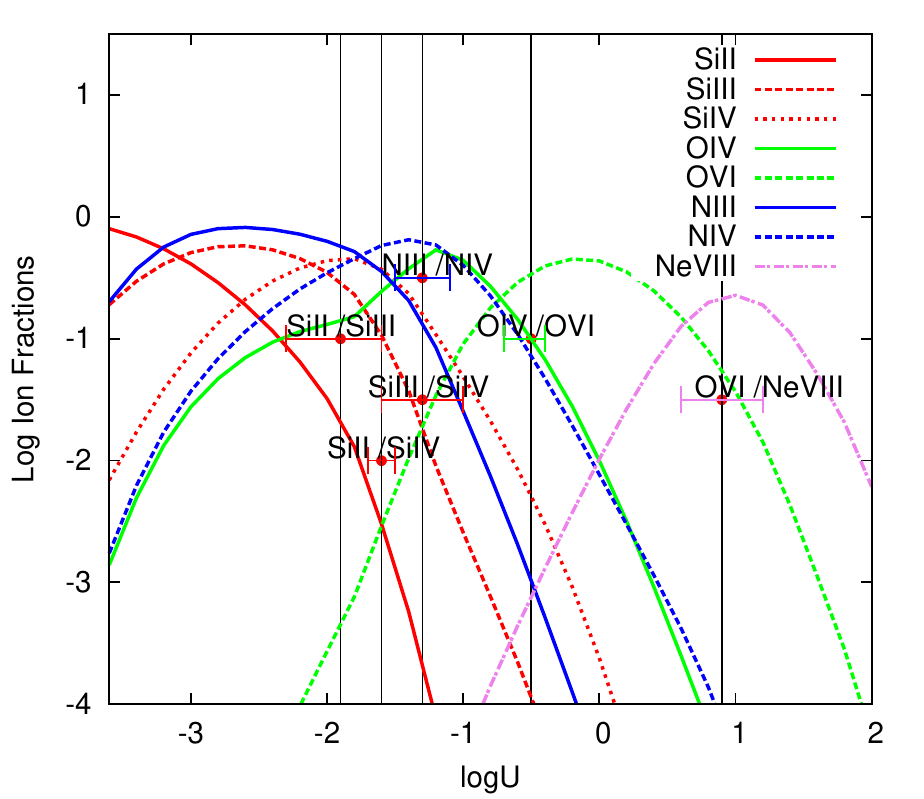}
\caption{Theoretical ionisation fractions, $f(\mbox{M}_i)$, for selected stages of the elements Si, N, O and Ne plotted against different ionisation parameters $\log U$ for component 2 in Q0119$-$046. The black vertical lines with error bars are the best estimations of $U$ for each ion pair.\label{fig:cloudy_1}}
\end{figure}

\begin{figure}
\centering
\includegraphics[width=84mm]{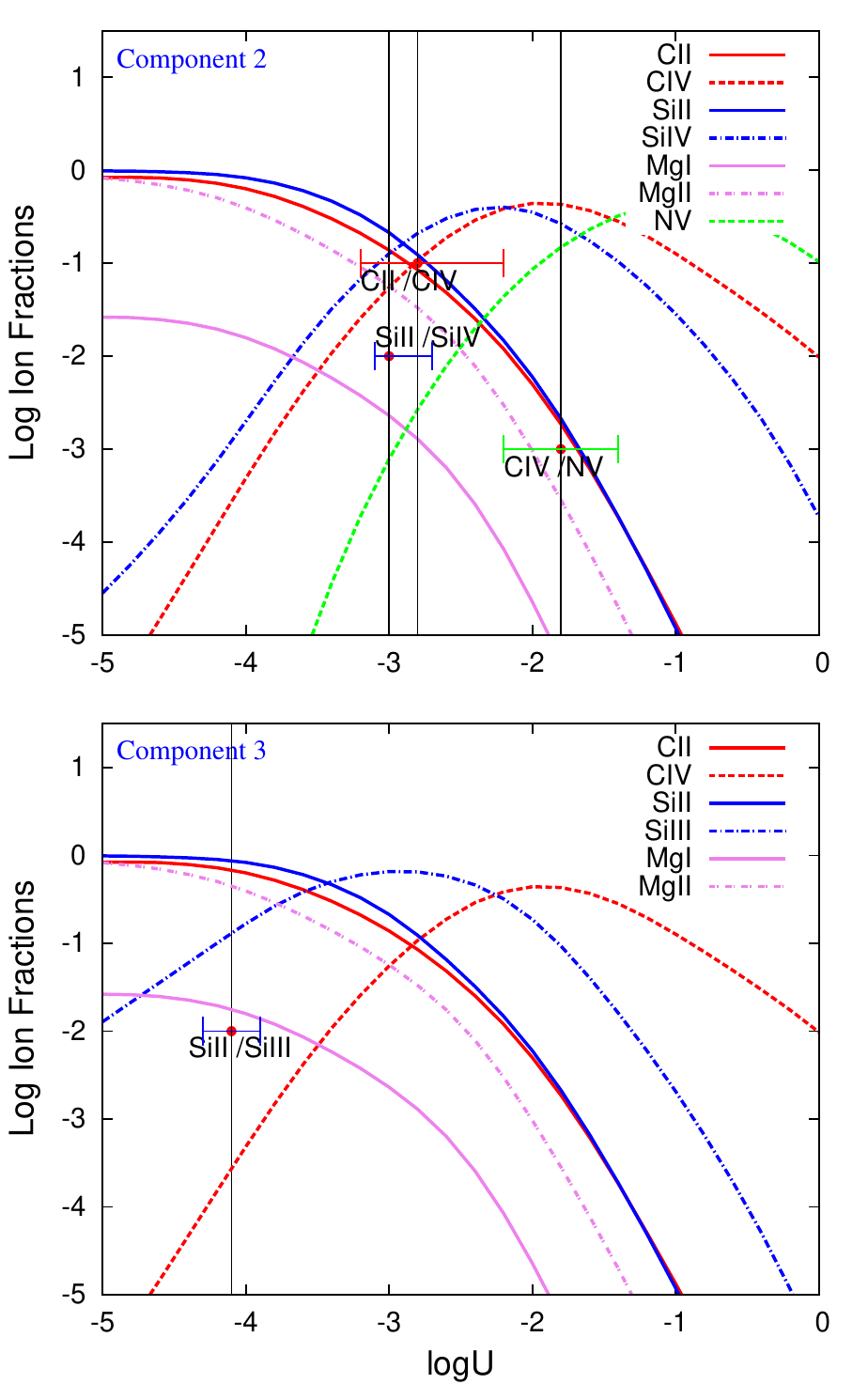}
\caption{Theoretical ionisation fractions, $f(\mbox{M}_i)$, for selected stages of the elements Si, C, Mg and N plotted against different ionisation parameters $\log U$ for components 2 and 3 in Q0105+061. The black vertical lines with error bars are the best estimations of $U$ for each ion pair.\label{fig:cloudy_2}}
\end{figure}

\begin{figure}
\centering
\includegraphics[width=84mm]{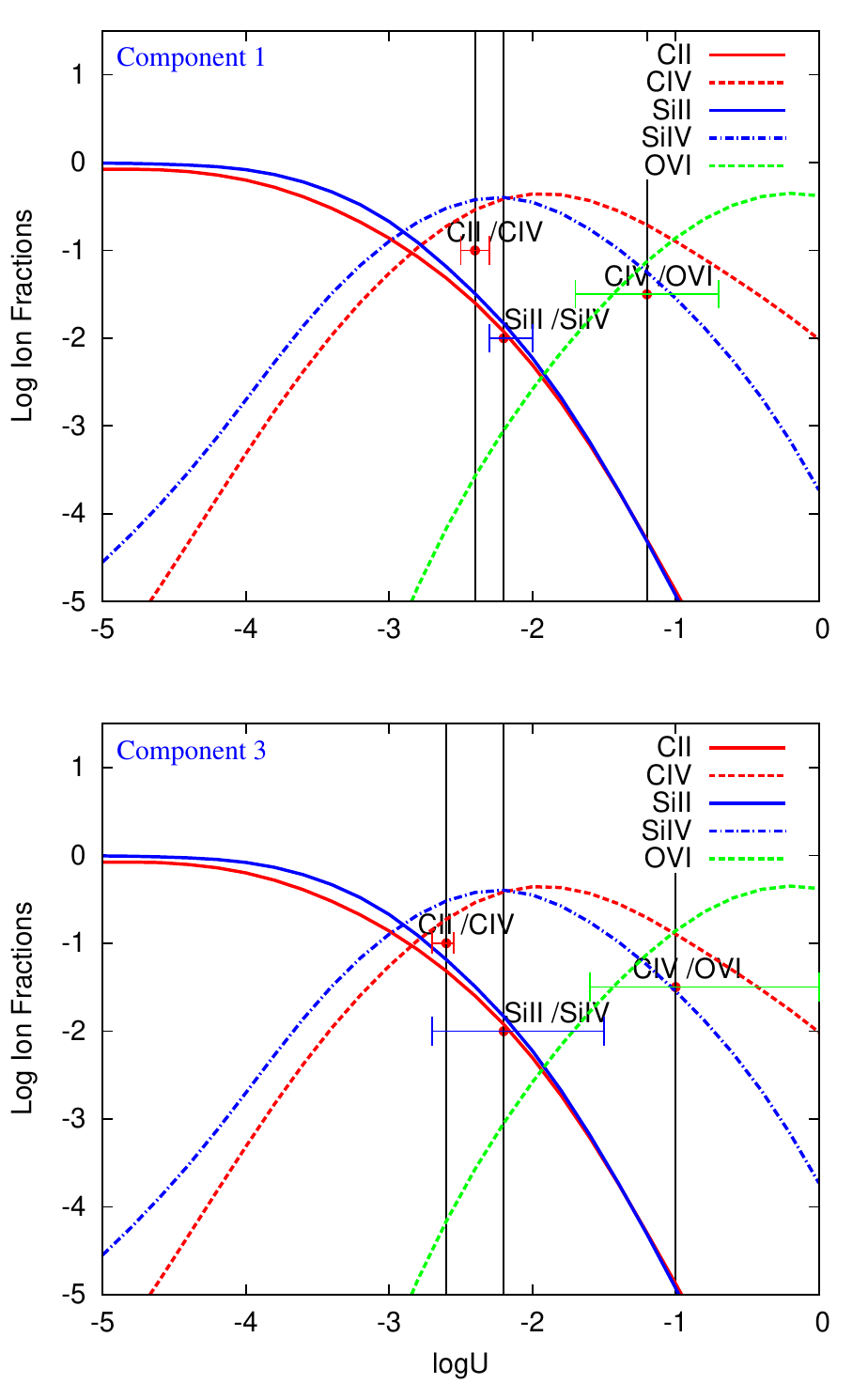}
\caption{Theoretical ionisation fractions, $f(\mbox{M}_i)$, for selected stages of the elements Si, C and O plotted against different ionisation parameters $\log U$ for components 1 and 3 in Q0334$-$204. The black vertical lines with error bars are the best estimations of $U$ for each ion pair.\label{fig:cloudy_3}}
\end{figure}

\begin{figure}
\centering
\includegraphics[width=84mm]{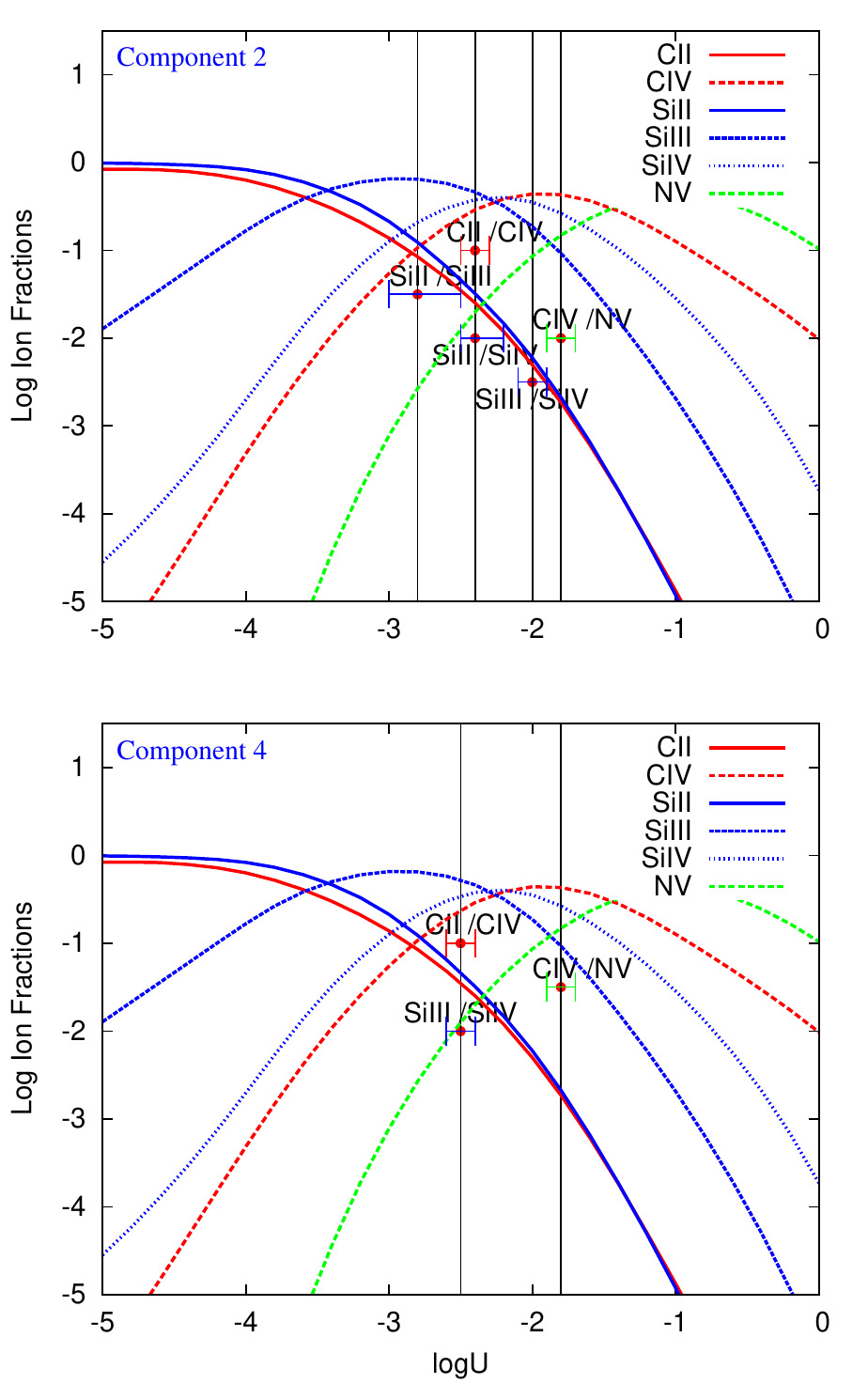}
\caption{Theoretical ionisation fractions, $f(\mbox{M}_i)$, for selected stages of the elements Si, C and N plotted against different ionisation parameters $\log U$ for components 2 and 4 in Q2044$-$168. The black vertical lines with error bars are the best estimations of $U$ for each ion pair.\label{fig:cloudy_4}}
\end{figure}

\Cref{fig:cloudy_1,fig:cloudy_2,fig:cloudy_3,fig:cloudy_4} show predicted ionisation fractions for the well-measured AAL systems we discuss below. These predictions compared to the measured column density ratios in various ion pairs (\cii/\civ , \siii/\siiv , \siii/\siiii , etc.) yield estimates of the ionisation parameter $U$ (shown by vertical lines in the figures) and its uncertainties (horizontal bars, based on the column density uncertainties in \Cref{tab:Q0119_1,tab:Q0119_2,tab:Q0105,tab:Q0334,tab:Q2044}). If the column density ratios of similar ion pairs (e.g. \cii/\civ\ and \siii/\siiv) are consistent with a single $U$ value, we derive a weighted mean $U$ and weighted error. The results are listed in \Cref{tab:para}. 

The temperature is the most important parameter in the far-UV. We ran additional \cloudy\ models with the temperatures 175,000 K and 700,000 K, which might be considered extreme ($\pm3 \sigma$) deviations from the continuum we adopted (Section 4.2). Generally, we found the ionisation parameter $\log U$ in the models changed by $\lesssim0.1$ if considering $1\sigma$ deviation, which is less than or similar to other uncertainties in the measured values of $\log U$.

Many of the systems in our sample exhibit a range of ions, e.g., from \mgi , \siii, \cii , and \mgii\ up to \nv , \ovi , and \neviii , that cannot coexist spatially in the same absorber. In principle, these observed lines could form in the same clouds at a single $U$ if the column densities are large enough to radiatively shield the lower ions behind a thick layer of ionised gas. However, in component 2 of Q0119$-$046, the amount of shielding is well constrained by our measurement of $N$(\hi ), e.g., the Lyman limit shown in \Cref{fig:HST,fig:limit_fit}. Our \cloudy\ simulations show that this produces minimal shielding with minimal effects on the ionisation structure. Therefore, a range in $U$ is required. 

The two $U$ values listed for Q0119$-$046 in \Cref{tab:para} illustrate the range. They derive from the column density ratio \siii/\siiii , which yields $\log U \sim$$-$1.9, up to \ovi/\neviii , which yields $\log U \sim$0.9 (assuming solar O/Ne abundances, see \Cref{fig:cloudy_1}). If the \neviii\ absorber is at the same distance from the quasar as the low-ionisation \siii -absorbing region, then the range in $U$ values implies that the \neviii\ gas is $\sim$630 times less dense than the \siii\ region (i.e., $n_e\sim$4 \cmn\ compared to $\sim$2500 \cmn\ for \siii). 

It is necessary to note that there is a large difference in the Doppler $b$ parameters between \siiii\ ($b\sim$64 \kms) and \siii\ ($b\sim$16 \kms). The large difference indicates that the lines sample different physical regions in the absorber or, perhaps, that the \siiii\ line is blended with a feature in the \lya\ forest. We note that the \siiii\ line width is similar to \siiv\ and, therefore, it seems likely that \siiii\ is broader than \siii\ because it has contributions from regions of higher ionization. 


We apply the same procedure to derive $U$ values and ranges for the other quasars, again assuming solar abundance ratio if ion pairs in the same elements are not available. \Cref{tab:para} lists the $U$ values and the ions used for each component. 

To estimate the total hydrogen column densities, we assume that the measured \hi\ column densities reside primarily with the lower metal ions (see Section 4.5 for more discussion). For example, for component 2 in Q0119$-$046, we apply an ionisation correction \hi /H~$\approx -3.2$ based on $\log U \sim$$-$1.9 from \siii /\siiii\ to the observed $N$(\hi ) value to derive a total column density of $\log N(\rm{H})$ (\cmN)~$=20.9$ in this absorber. For the other quasars with only conservative lower limits on $N$(\hi ), we derive conservative lower limits on $N$(H). These results are also listed in \Cref{tab:para}. 

It is important to note that the results for $U$ and $N(H)$ for the other quasars, with only lower limits on $N$(\hi ), are based on \cloudy\ models where the absorber is optically thin in the ionising continuum (Section 4.2). If shielding does play a role, then the total column densities would need to be at least as large as Q0119$-$046, where $\log U \sim$$-$1.9 and $\log N(\rm{H})$ (\cmN)$\;=20.9$ combined to produce significant absorption at the \hi\ Lyman edge. We cannot exclude this possibility for the other quasars, but their absorbers are clearly different from Q0119$-$046 given their lower densities and larger distances (discussed below). 

Another constraint on the ionisation and densities comes from the neutral \mgi\ \lam 2853 line that appears uniquely in components 2 and 3 of Q0105+061. This ion cannot be shielded behind an \hii --\hi\ recombination front because its ionisation potential 7.65 eV is well below that of \hi\ at 13.6 eV. Therefore, the \mgi\ lines suggest that high densities lead to very low ionisation parameters in some portions of these absorbers. Our \cloudy\ simulations (\Cref{fig:cloudy_2}) indicate that significant amounts of \mgi\ require $\log U<-5$. (The results are the same if we use larger column densities like Q0119$-$046 because the shielding effects for \mgi\ are still negligible.) For component 2 in Q0105+061 with estimated $\log U\approx-3.0$ in the \siii\ region (top panel of \Cref{fig:cloudy_2}), we estimated an upper limit on the density of $n_e\lesssim 7$ \cmn\ (\Cref{tab:para}). If the \mgi\ line forms at the same radial distance as \siii , then there would need to be regions with densities $\gtrsim$100 times larger than the \siii\ gas to support measurable \mgi .

\begin{table*}
\centering
\caption{Parameters of some absorbers. Columns show quasar name, bolometric luminosity in $10^{47}$ \erg, component, velocity shifts in \kms, column density ratio of \cii* and \cii, column density ratio of \siii* and \siii, electron density, ionisation parameter, total hydrogen column density, and radial distance in kpc.}
\label{tab:para}
\tabcolsep=0.15cm
\begin{tabular}{@{}cccccccccc}
\hline \hline
QSO & $L_{Bol}$ & Component & $v$ & $\frac{N(\cii*)}{N(\cii)}$ & $\frac{N(\siii*)}{N(\siii)}$ & $\log n_e$ & $\log U$ & $\log N(\mbox{H})$ & $R$ \\
  & ($10^{47}$ \erg) &  & (\kms) &  &  & (\cmn) &  & (\cmN) & (kpc) \\
\hline
Q0119$-$046 & 4.2 & 2 & $71\pm202$ & $1.9\pm 0.2$ & $1.2\pm 0.3$ & $3.4\pm 0.3$ & $-1.9\pm 0.3$ (\siii\ and \siiii)  & $20.9\pm 0.4$  & $5.7^{+5.6}_{-2.8}$ \\ [-2pt]
& & & & & & & $0.9\pm0.3$ (\ovi\ and \neviii) & & \\
\hline
Q0105+061 & 2.9 & 1 & $-2894\pm507$ & $<0.2$ & $<0.5$ & $<0.7$ & $-2.8\pm0.3$ (Si and C ions) & $>15.9$ & $>300$ \\ [6pt]
& & 2 & $-2858\pm507$ & $<0.01$ & $<0.01$ & $<0.9$ & $<-5$ (\mgi) &$>15.9$ & $>300$ \\  [-2pt]
&  & & & & & & $-3.0\pm 0.2$ (Si and C ions) & & \\  [-2pt]
&  & & & & & & $-1.8\pm0.4$ (\civ\ and \nv) & & \\ [6pt] 
 & & 3 & $-2797\pm507$ & $<0.05$ & $<0.1$ & $<1.9$ & $<-5$ (\mgi)& $>14.5$ & $>331$ \\  [-2pt]
&    & & & & & &  $-4.1\pm 0.2$ (\siii\ and \siiii) & & \\ [6pt]
 & & 4 & $-2590\pm507$ & $<0.3$ & $<0.6$ & $<0.9$ & $-2.9\pm0.2$ (Si and C ions) & $>15.7$ & $>267$\\ [6pt]
&   & 5 & $-2544\pm507$ & $<0.03$ & $<0.2$ & $<2.2$ & $-2.6\pm 0.1$ (Si ions) & $>16.6$ & $>41$\\ [6pt]
 &  & 6 & $-2503\pm507$ & $<0.01$ & $<0.03$ & $<1.4$ & $-2.6\pm 0.3$ (Si ions) & $>15.8$ & $>109$\\
\hline
Q0334$-$204 & 3.4 & 1 & $-3035\pm94$ & $<0.5$ & $<0.2$ & $<1.2$ & $-2.2\pm0.1$ (Si and C ions) & $>18.4$ & $>50$\\ [-2pt]
 &   & & & & & &  $-1.2\pm0.5$ (\civ\ and \ovi) & & \\ [6pt]
 &  & 2 & $-3006\pm94$  & $<0.2$ & $<0.2$ & $<0.7$ & $-2.3\pm0.3$ (Si and C ions) & $>18.4$ & $>99$\\ [-2pt]
 &   & & & & & &  $-1.3\pm0.7$ (\civ\ and \ovi) & & \\ [6pt]
 &  & 3 & $-2991\pm94$  & $<0.2$ & $<0.3$ & $<0.7$ & $-2.6\pm0.1$ (Si and C ions) & $>18.0$ & $>140$\\ [-2pt]
  &    & & & & & &  $-1.0^{+1.0}_{-0.6}$ (\civ\ and \ovi) & & \\ 
\hline
Q2044$-$168 & 3.5 & 1 & $-2113\pm102$ & $<0.5$ & --- & $<1.2$ & $-2.0\pm0.2$ (Si and C ions) & $>17.0$ & $>39$\\ [6pt]
 &  & 2 & $-2042\pm102$ & $<0.07$ & $<0.08$ & $<0.2$ & $-2.4\pm0.1$ (Si and C ions) & $>16.8$ & $>198$\\ [-2pt]
&    & & & & & & $-1.8\pm0.1$ (\civ\ and \nv) & & \\ [6pt] 
&   & 3 & $-1960\pm102$ & $<0.2$ & --- & $<0.7$ & $-2.7\pm0.2$ (Si and C ions) & $>16.0$ & $>157$\\ [6pt]
&   & 4 & $-1929\pm102$ & $<0.1$ & --- & $<0.4$ & $-2.5\pm0.1$ (Si and C ions) & $>16.6$ & $>176$\\ [-2pt]
&    & & & & & & $-1.8\pm0.1$ (\civ\ and \nv) & & \\ [6pt] 
&   & 5 & $-1878\pm102$ & $<0.3$ & --- & $<0.9$ & $-2.3\pm0.1$ (Si and C ions) & $>16.7$ & $>79$\\
\hline
\end{tabular}
\end{table*}

\subsection{Radial Distance}

Here we estimate the radial distance, $R$, of the \siii\ or \cii\ AAL regions from the quasars using $n_e$ and $U$ in the \siii\ or \cii\ regions derived in Sections 4.1 and 4.3, respectively. First we combine our adopted quasar spectrum from Section 4.2 with \Cref{eq:8} and \Cref{eq:9} to derive an expression for the distance,
\begin{equation}
\label{eq:10}
R=40.8\left(\frac{\nu L_{\nu}(1500 \rm{A})}{10^{46}\ \rm{ergs/s}}\right)^{\frac{1}{2}}\left(\frac{10\ \rm{cm}^{-3}}{n_{\rm{H}}}\right)^{\frac{1}{2}}\left(\frac{0.01}{U}\right)^{\frac{1}{2}}\ \rm{kpc},
\end{equation}
where $\nu L_{\nu}(1500 \rm{A})$ is the monochromatic luminosity at 1500 \AA , and we assume $n_e\approx n_H$ for an ionised gas. We estimate luminosities for each quasar using the $g$-band photometry from the Sloan Digital Sky Survey (SDSS). This provides a flux at rest wavelengths near 1500 \AA, which we then extrapolate to 1500 \AA\ using a power law with spectral index $\alpha_{OX}=-1.7$. We estimate the bolometric luminosity, $L_{Bol}$, for each quasar via integration of the SED, $\sim6.76\nu L_{\nu}(1500 \rm{A})$ (\Cref{tab:para}). For Q0119$-$046, the derived density and $\log U \sim$$-$1.9 appropriate for the \cii\ and \siii\ absorbing region yields a distance of $R\sim$5.7 kpc (see \Cref{tab:para}). The errors listed for this distance derive from the uncertainty in $U$. For the other quasars, we combine the density upper limits with $\log U$ listed in \Cref{tab:para} to derive conservatively small distance lower limits that range from $R\gtrsim 40$ to 330 kpc. 

\subsection{Gas Metallicity}

We estimate metallicities only for component 2 in Q0119$-$046 where $N(\hi)$ is well measured. (The lower limits on $N(\hi)$ in the other quasars lead to large upper limits on the metallicities, well above solar, that do not provide useful constraints.) The metallicity can be written generally as
\begin{equation}
\label{eq:11}
\left[\frac{\mbox{M}}{\mbox{H}}\right]=\log\left(\frac{N(\mbox{M}_i)}{N(\hi)}\right)+\log\left(\frac{f(\hi)}{f(\mbox{M}_i)}\right)+\log\left(\frac{\mbox{H}}{\mbox{M}}\right)_{\odot},
\end{equation}
where $(\mbox{H/M})_{\odot}$ is the solar abundance ratio of hydrogen to some metal $\mbox{M}$, $N(\hi)$ and $f(\hi)$ are the column density and ionisation fraction in \hi, respectively, and $N(\mbox{M}_i)$ and $f(\mbox{M}_i)$ are the column density and ionisation fraction in some ion $\mbox{M}_i$ of metal $\rm{M}$. 

To calculate [Si/H] and [C/H] for this absorber, we first use $\log U\approx-1.9$ obtained from the \siii\ and \siiii\ region to determine the ionisation fractions $f(\cii)$, $f(\siii)$, and $f(\hi)$ from our calculation in \Cref{fig:cloudy_1}. Then the measured values of $N(\siii)$ and $N(\cii)$ combined with $N(\hi)$ indicate [C/H]~$\approx -1.8\pm 0.1$ and [Si/H]~$\approx -2.4\pm 0.2$. From Section 4.3, we know there is a range of ionization for this component. If assuming the high-ions \ovi\ and \neviii\ have similar metallicity as the low-ions \siii\ and \cii, then we use the $\log U\sim$0.9 from the ratios \ovi/\neviii, and the corresponding $f(\hi)\sim$$-$6.7 and $f(\ovi)\sim$$-$1.4 to predict $N$(\hi). The predicted $\log N$(\hi) is $\sim$15.5, which is much less than $\sim$17.7. This is the reason why we assume that the measured \hi\ column densities reside primarily with the lower metal ions (Section 4.3). 

\subsection{Spatial Structure and Cloud Survival}

There is partial covering in roughly half of the components in Q0119$-$046. This implies that absorbers are not much larger (and probably smaller) than the projected area of the emission regions. The partial covering in \civ\ and \lya\ appears to apply to the broad emission line region (BLR) because 1) these absorption lines sit on top of strong broad emission lines \citep{Sargent82}, 2) the covering fraction in \civ\ is only slightly less than \siiv, which does not sit on a strong emission line, and 3) the bottoms of the \lya\ troughs are slanted in a way that is consistent with the peak of the broad \lya\ emission line partially filling in the troughs. The BLR scaling relationship in \citep{Bentz13} indicate that the size of the broad emission region is $\sim$1.0 pc, and therefore the absorbers have transverse size $\lesssim$1.0 pc. We also find evidence for partial covering in \ovi, \neviii\ and the higher Lyman lines. Since these features do not sit on strong emission lines, they must partially cover the much smaller continuum source. This requires the absorber sizes that are conservatively less than 0.01 pc \citep[][Hamann et al., submitted]{Netzer92, Hamann11}. 

These small cloud sizes in Q0119$-$046 are surprising given the derived radial distance of $\sim$5.7 kpc, i.e., far for the quasar in the host galaxy. Such clouds are not expected in a normal galactic interstellar medium, but they not unprecedented for distant AAL absorbing regions. For example, \cite{Hamann01} found partial covering of the continuum source in an AAL absorber $\sim$28 kpc from the central quasar. If the clouds are not confined by an external pressure, they will dissipate in roughly a sound crossing time \citep{Hamann01, Finn14} given by 
\begin{equation}
t_{sc}=\frac{l}{c_s},
\end{equation}
where $c_s$ is the sound speed and $l$ is the characteristic cloud size \citep{Schaye01, Hamann01, Finn14}. For a nominal temperature of $10^4$ K and $l \lesssim 0.01$ pc or more conservatively $l \lesssim 1$ pc, the cloud survival times are $\lesssim$700 yr or $\lesssim$70,000 yr, respectively. The gas speeds in the various AAL systems in Q0119$-$046 are overall $\lesssim$1200 km~s$^{-1}$, which means that the cloud survival times are much shorter than any reasonable flow time we might assign to these absorbers. It therefore seems likely that the clouds were created in situ, at or near their observed location $\sim$5.7 kpc from the quasar (see Section 5 for more discussion). 


\section{Discussion}

The information derived in Sections 3 and 4 provides valuable constraints on the nature and origins of the AALs in our quasar sample. The results might have general relevance to high-redshift quasar environments, but it is important to keep in mind that the sample is biased. The quasars were selected to have rare low-ionisation lines (\siii\ and/or \cii ) useful for density and location constraints. More common types of AALs, with only higher-ionization lines, are not included in our sample. The full range of ions detected in our study, from \mgi, \mgii, \siii, \cii, up to \civ, \nv, and in some cases \ovi\ and \neviii, often require a range of ionisation parameters in the AAL gas. If the diverse lines in each system form at roughly the same location, then the absorbing regions must span a range of densities. The most extreme case is component 2 in Q0119$-$046, where ions ranging from \siii\ to \neviii\ indicate a factor of $\sim$630 range in densities (Section 4.3). This result is similar to several other quasars where measured \neviii\ AALs also indicate a range of ionisations and densities \citep{Petitjean99, Hamann00, Hamann95, Arav13, Muzahid13, Finn14}.

In our study, Q0119$-$046 has a unique dataset because we combine high-resolution spectra from Keck-HIRES with spectra from HST-FOS that reach wavelengths down to $\sim$750 \AA\ in the quasar rest frame. These data provide the best measurements and different results than the other three AAL quasars in our sample. Thus we discuss Q0119$-$046 separately below. 

\subsection{Q0119$-$046}

The AALs in Q0119$-$046 are clearly intrinsic to the quasar environment based on multiple components with partial covering of the background light source and high densities of $n_{\textrm{H}}\sim 2500$ \cmn\ that lead to a derived distance of $R\sim$5.7 kpc. These specific density and location results are similar to the absorbers studied by \citet{Dunn10, Arav13, Finn14}, which have $n_H \sim 1000$ to 6000 cm$^{-3}$ and $R\sim$2 to 6 kpc. The partial covering we find in Q0119$-$046 additionally implies that the absorber is composed of small clouds with characteristic sizes $\lesssim$1 pc and possibly $\lesssim$0.01 pc (if the partial covering applies to the continuum source, Section 4.6). Our results overall are consistent with the previous study of Q0119$-$046 by \cite{Sargent82} that obtained only a lower limit on the density from \cii */\cii\ and an upper limit on the distance of $<$60 kpc. The AALs in Q0119$-$046 join a growing number of well-measured systems in other quasars where density indicators place the absorbers within the quasar environments but at large distances of a few to a few hundred kpc, and where partial covering of the quasar emission source can require small cloud sizes at these distances \citep[see also][]{Tripp96, Barlow97, Srianand00, Hamann01, Gabel06, Arav08}.

The radial distance $R\sim$5.7 kpc of the absorber in Q0119$-$046 is an interesting location where a quasar-driven wind might be interacting with interstellar gas in the extended host galaxy. The derived metallicities in the range $\sim$0.004 to $\sim$0.016 $Z_o$ are substantially smaller than expected for an outflow originating in the galactic nuclear regions near the quasar, where solar or higher metallicities should be present \citep{Hamann99, Arav01, Hamann02, Dietrich03, Warner04, Gabel05b, Nagao06, Simon10}. If this AAL complex stems from a quasar-driven outflow, then the low metallicities might imply that the outflow gas is mixed with ambient interstellar gas in the galaxy. The multi-component nature of the AAL complex might identify interstellar clouds that have been shredded and dispersed by an unseen high-speed quasar-driven outflow, as described in some recent theoretical models \citep{Hopkins10, Faucher11, Faucher12b}. This interpretation is appealing because it provides a natural mechanism for creating small absorption-line clouds in situ, and thus avoiding the problem of their short survival times (see refs above, also Section 4.6).

However, another possibility is that the AAL absorbers in Q0119$-$046 represent infalling metal-poor gas from the intergalactic medium (IGM). The measured line velocities favor this infall interpretation. Components 1 and 2 have velocity shifts consistent with zero, while the other systems span a range of {\it positive} velocities from $\sim$240 to $\sim$1150 \kms\ (\Cref{tab:Q0119_1} and Section 2.1). In this infall scenario, small absorption-line clouds could be created in situ as condensations in larger reservoir of cold-mode accreting gas \citep{Keres09, Fumagalli11, Hafen16}. \citet{McCourt16} describe theoretically that clouds of optically-thin, pressure-confined gas are inclined to fragment as they cool, reaching very small size scales of $\sim$0.1 pc. 

Accurate AAL velocity shifts are critical for this analysis. We adopt a redshift for Q0119$-$046 from \citet{Steidel91} based on the \mgii\ broad emission line, with estimated measurement uncertainties of $\lesssim100$ \kms\ (Section 2.1). Another possible source of uncertainty is the offset of the \mgii\ line from the true systemic rest frame of the quasar environment. Studies of large quasar samples indicate that the mean offset of the \mgii\ emission line relative to [\oiii] $\lambda$5007 corresponds to a slight blueshift of $\sim$100 \kms\ with 1 $\sigma$ scatter of $\sim$270 \kms\ \citep{Richards02, Shen07, Shen16}. If the [OIII] line is a better redshift indicator, the probability for a random \mgii\ blueshifted matching our measured shift of $\sim$1150 \kms\ is less than 0.005\%\ ($\sim$4$\sigma$). We conclude that at least some of the AAL components in Q0119$-$046 form in gas that is infalling toward the quasar. 

Infalling gas (e.g., cold mode accretion) from the IGM is believed to be important during the early stages of galaxy formation to build mass, trigger star formation, and fuel the central black hole \citep{Katz03, Keres09, Keres12}. It is likely that infall and outflow occur together if cold-mode accretion is involved in triggering the starbursts and quasars that also drive feedback \citep{Costa14, Nelson15, Suresh15}. Recent observations show that massive gas reservoirs are indeed present around high-redshift quasars, and that they are more extended and more massive around quasar hosts than similar inactive (non-quasar) galaxies \citep[e.g.,][]{Prochaska14, Johnson15, Martin15, Martin16, Borisova16, Bouche16, Ho17}. The nature of these gas reservoirs is poorly understood, but they are consistent with enhanced infall/cold-mode accretion from the IGM during an early active stage of massive galaxy evolution when there is ongoing quasar activity. Quasar AALs can be valuable tracers of infall because they measure the gas speeds and physical conditions along radial lines of sight into galactic nuclei. The AALs in Q0119$-$046 might provide direct observational evidence for infall related to the assembly of a massive galaxy at redshift $z\sim$1.96.

Compared to the work by \citet{Sargent82}, we obtain more accurate results, especially $N(\hi)$ and $N(\rm{H})$, based on the high-resolution Keck spectrum and more lines including the Lyman series and the Lyman limit from the HST spectrum. We estimate $n_{\rm{H}}=10^{3.4\pm0.3}$ \cmn\ via the strength ratio of \siii*/\siii. While they constrained $n(\rm{H})>100$ \cmn\ based on the existence of excited state \cii*. The radial distance we obtain is $\sim$5.7 kpc because of our accurate $n_{\rm{H}}$ estimates. While they estimated its radial distance to be less than 60 kpc by assuming $n_{\rm{H}}>100$ cm$^{-3}$ and $N(\rm{H})=10^{20}$ cm$^{-2}$. We find that $\sim$50\%\ of the absorbers show partial covering cases based on the lines \civ, \siiv, \ovi, \neviii, Lyman series, and the Lyman limit. This requires absorber size scales less than 0.01 pc. While they did not find the partial covering cases. We find that the gas clouds are metal-poor and infalling at velocities from 0 to 1100 \kms\ by using the emission-line redshift obtained from \mgii. While they use the emission line \civ\ to obtain the systemic redshift, which leads to a smaller emission-line redshift than the real one, and thus they obtain a very large infalling speed of 2800 \kms\ for the absorber at $z_{abs}=1.9644$. They guessed that the absorbers are analogous to the NGC 1275 high-velocity filaments \citep{Kent79}. From our results, we conclude, because of these metal-poor, tiny ($<0.01$ pc) absorbers and their short survival time ($<700$ yr), they are created in situ, probably as condensations in cold-mode accreting gas or shredded IGM clouds that are dispersed by an unseen high-speed quasar-driven outflow. In addition, there seems to be no variation between the spectra of the two periods by visual comparisons, if the resolution difference is considered. The lack of variability is consistent with the absorber residing at kpc distances. 

\subsection{Q0105+061, Q0334$-$204, and Q2044$-$168}

The AALs in the other three quasars in our sample, Q0105+061, Q0334$-$204, and Q2044$-$168, have velocity blueshifts consistent with outflows at speeds of $-$1900 to $-$3000 \kms . These systems do not exhibit partial covering of the background light source, and the upper limits we derive on their densities, $\lesssim$150 to 15 \cmn, place the absorbers at large distances from the quasars, namely, $\gtrsim$40 to $\gtrsim$330 kpc. Thus the relationship of the AAL systems to the quasars is not known. It is possible that they form in intervening gas that is not physically related to the quasars, such as galactic halos in the same galaxy cluster or group as the quasar host galaxies. However, previous statistical studies of large quasar samples indicate that $\sim$80 percent of \civ\ AALs with rest equivalent width REW$\geq$0.3 \AA\ at these velocity shifts form in quasar-driven outflows \citep{Misawa07, Nestor08, Wild08, Simon10}. 

If the AALs we measure do identify quasar outflows, then their speeds and minimum radial distances yield estimates of the minimum outflow masses and kinetic energies. We estimate the total masses assuming the absorbers are part of a spherical shell at distance $R$ from the central quasar, namely
\begin{equation}
M=4\pi\mu m_pQR^2N(\rm{H}),
\end{equation}
where $Q$ is the global covering factor of the absorber as seen from the central quasar, $m_p$ is the mass of a proton, and $\mu\sim$1.4 is the mean molecular weight per proton in an ionized plasma with solar abundances \citep{Hamann00iop, Hamann01, Dunn10}. A rough estimate of $Q$ is the incidence of associated absorbers generally in quasar spectra. Previous studies show that the incidence of AALs in radio-quiet quasars is around 20\% \citep{Nestor08}. Therefore, we adopt $Q=0.2$ to derive minimum total masses of $M\gtrsim (0.1-7.0)\times10^7$ \msun\ for the AAL absorbers in the three quasars. The corresponding minimum kinetic energies, given by $K=M\textrm{v}^2/2$ are quite small, in the range $\sim$$(2.5-100)\times 10^{48}$ \erg. Dividing these energies by a characteristic flow time, $\Delta t\sim R/$v, yields kinetic energy rates, $\dot{K}$, that we compare to the bolometric luminosities (Section 4.4 and \Cref{tab:para}) to derive minimum ratios $\dot{K}/L_{Bol} \gtrsim 10^{-7}$ to $10^{-5}$. The lower limits on these ratios are much too small to be important for feedback to the host galaxies, where $\dot{K}/L_{Bol} \gtrsim 0.05$ to 0.005 is believed to be required \citep{Scannapieco04, Hopkins10}. However, these estimates are based on conservative lower limits on $N(H)$, which are 2 to 4 dex smaller than our reliable $N(H)$ measurement for component 2 in Q0119$-$046. 

We conclude that the AAL systems in these three quasars are likely to form in quasar-driven outflows, but their energies and potential for feedback to the host galaxies cannot be determined from existing data. 


\section{Summary}

We discuss rest-frame UV spectra of four redshift 2--3 quasars, Q0119$-$046, Q0105+061, Q0334$-$204, and Q2044$-$168, selected to have low-ionisation AALs in \siii\ and/or \cii\ that are valuable to estimate electron densities and radial distances from the quasars. The data include high-resolution spectra obtained with Keck-HIRES or VLT-UVES that we combine, for Q0119$-$046 only, with an HST-FOS spectrum that reaches down to $\sim$750 \AA\ in the quasar rest frame. Our analysis of Q0119$-$046 builds upon previous work on by \cite{Sargent82}, but with higher-quality data and wavelength coverage that measures below the \hi\ Lyman limit. The other three quasars were not previously studied. We fit every detected AAL to measure the gas kinematics, column densities, and line-of-sight covering fractions, and we estimate ionisations, total column densities, and radial distances using \cloudy\ photoionisation models.

Our main results for Q0119$-$046 are the following: 

1) The AALs in Q0119$-$046 identify a complex absorbing structure with at least 11 distinct velocity components. The velocity shifts of these components, ranging from $\sim$0 to roughly +1150 \kms\ (relative to the \mgii\ emission line), are indicative of infall towards the quasar with estimated uncertainties of $\lesssim$200 \kms .

2) The electron density implied by the \siii*/\siii\ line ratios in component 2, $n_e \sim 2500$ \cmn\  (Section 4.1), indicates that this absorber resides at a distance $\sim$5.7 kpc from the quasar (Section 4.4). 

3) The wide range of ions detected in these AALs, from \siii\ up to \neviii , cannot be attributed to radiative shielding effects inside the absorber. It requires a range of densities from $\sim$2500 \cmn\ in the \siii\ region down to $\sim$4 \cmn\ in the \neviii\ gas {\it if} the lines all form at roughly the same radial distance (Sections 4.3 and 5.1).

4) The metallicities in AAL component 2 of Q0119$-$046 are in the range $\sim$0.004 to $\sim$0.016 times solar (Section 4.5). 

5) Roughly half of the AAL components in Q0119$-$046 partially cover the background emission source (Section 4.6). This implies that the absorbers are composed of small clouds with characteristic sizes $\lesssim$1 pc and possibly $\lesssim$0.01 pc (based on \ovi, \neviii\ and the higher Lyman lines that partially cover the quasar continuum source). 

6) These tiny AAL clouds will have short survival times of $\lesssim$700 yr or $\lesssim$70,000 yr if they are not confined by an external pressure. At the derived distance of $\sim$5.7 kpc, the cloud survival times are much less than a flow time, suggesting that the clouds are created in situ, at their observed location. 

7) These results for Q0119$-$046 are consistent with models of galactic interstellar clouds being shredded and dispersed by a quasar-driven wind. However, the evidence for infall in the line shifts strongly favors an interpretation of this AAL complex as a series of condensations (spanning a factor of $\sim$630 in density) embedded in a medium that is cold-mode accreting from the intergalactic medium (Section 5.1). 

The other three quasars in our sample, Q0105+061, Q0334$-$204, Q2044$-$168, have AAL properties similar to each other but different from Q0119$-$046. Our main results for those quasars are the following: 

1) Non-detections of the excited-state \cii* and \siii* lines yield upper limits of electron densities from $n_e \lesssim150$ to $\lesssim$15 \cmn\ and lower limits on the radial distances from $R\gtrsim 40$ to $\gtrsim$330 kpc (Sections 4.1 and 4.4). 

2) Most of the components in these AAL systems exhibit a range of ionizations, including the neutral \mgi\ $\lambda$2853 line in Q0105+061. These ranges in ionization again require a range of densities within individual velocity components if the lines all form in roughly the same location (Section 4.3).

3) There is no evidence for partial covering in any of these AAL systems. Deep saturated lines in some of the components clearly indicate covering fractions of unity. 

4) The AAL velocity shifts are indicative of outflows at speeds of $\sim$1900 to $\sim$3000 \kms. Previous studies of large quasar samples indicate that $\sim$80 percent of strong AALs in this velocity range do form in quasar outflows. These AALs might represent highly extended quasar-driven outflows, although the physical relationship of these particular AALs to the quasars cannot be determined (Section 5.2). 

\section*{Acknowledgments}

We thank the anonymous referee for useful comments and suggestions. This work was supported by University of California, Riverside.

\bibliographystyle{mnras}
\bibliography{reference}




\label{lastpage}

\end{document}